\documentclass[twocolumn]{aastex701}

\usepackage{amsmath} 
\usepackage[dvipsnames]{xcolor}

\usepackage{subcaption}
\usepackage{units}
\usepackage{bm}
\usepackage{hyperref}

\begin{document}

\title{A comprehensive framework for phase-coherent mapping of the gravitational-wave sky \\ with pulsar timing arrays}

\author[orcid=0000-0002-7031-4828,sname='M. Cury\l{}o']{Ma\l{}gorzata Cury\l{}o}
\affiliation{School of Physics and Astronomy, Monash University, Clayton VIC 3800, Australia}
\affiliation{OzGrav: The ARC Centre of Excellence for Gravitational Wave Discovery}
\email[show]{gosia.curylo@monash.edu}  

\author[orcid=0000-0002-4418-3895,sname='E.Thrane']{Eric Thrane} 
\affiliation{School of Physics and Astronomy, Monash University, Clayton VIC 3800, Australia}
\affiliation{OzGrav: The ARC Centre of Excellence for Gravitational Wave Discovery}
\email[]{eric.thrane@monash.edu}  

\author[orcid=0000-0003-3763-1386, sname='P. D. Lasky']{Paul D. Lasky} 
\affiliation{School of Physics and Astronomy, Monash University, Clayton VIC 3800, Australia}
\affiliation{OzGrav: The ARC Centre of Excellence for Gravitational Wave Discovery}
\email[]{paul.lasky@monash.edu}  

\author[sname='Dawson Gaynor']{Dawson S Gaynor} 
\affiliation{Department of Physics and Astronomy, Carthage College, Kenosha, WI 53140 USA}
\email[]{dawsongaynor04@gmail.com}


\begin{abstract}
We present a practical implementation of a phase-coherent mapping technique for pulsar timing arrays that resolves the full complex polarisation state of the gravitational-wave sky as a function of direction and frequency. Unlike standard cross-correlation methods, this approach preserves the amplitude, phase, and polarisation of the signal in every sky pixel. The resulting maps constitute a compact, minimally processed summary of the data from which all subsequent analyses --- characterisation of a stochastic background, searches for anisotropy, and identification of individual sources --- can be derived within a single unified framework. Our implementation is fully compatible with established pulsar timing data analysis methods. We validate the framework through a series of realistic simulations with varying array configurations, noise properties, and signal types. We demonstrate robust recovery of source amplitudes and sky locations across different scenarios, and discuss the impact of polarisation leakage, noise, and direction-dependent array sensitivity on the recovery of astrophysical signals. 
\end{abstract}



\section{Introduction}\label{sec:intro}
A robust detection of nanohertz gravitational waves (GWs) will have immense consequences for our understanding of the evolution of galaxies and their central supermassive black holes (SMBHs). It will provide the missing direct evidence that SMBHs can form close sub-parsec binaries and merge, and will carry a wealth of information about galactic interiors. 

Pulsar timing arrays (PTAs) are designed to search for nanohertz-frequency GWs by means of identifying characteristic deviations in the arrival times of radio pulses from millisecond pulsars. In fact, six collaborations, spread around the world---including the Chinese PTA, European PTA, Indian PTA, NANOGrav, the MeerKAT PTA and the Parkes PTA---have put forward the very first evidence of the GW background emerging from their timing data \citep{CPTA_GWB,PPTA_GWB,EPTA_GWB,NG_GWB,MK_GWB}. The result is extremely encouraging. Nonetheless, its significance is still too low to confidently infer the signal's properties (e.g., its spectral shape and origin).
This is likely to change soon due to ongoing efforts in collection of new data, the combination of data from multiple PTAs, increasing sensitivity of the instruments, and improved analysis techniques (e.g. \citealp{Curylo25, Grunthal25, Wright25, Mingarelli26}, Di Marco et. al., in prep.). 

The most likely source of low-frequency GWs are SMBH binaries.
However other, more exotic scenarios involving, e.g., cosmic strings or phase transitions in the early Universe, are also possible (see e.g., \citealp{Burke19} and references therein). One of the commonly explored avenues to distinguish between them is to examine the signal's spectral properties. However, this approach is subject to important caveats, mainly because of a broad range of theoretically predicted spectra both in the case of poorly constrained cosmological models as well as SMBH binary emission which is affected by environmental effects. For example, in the simplest scenario of circular SMBH binaries whose inspiral is driven by GWs only, the associated spectral index of the GW background is predicted to be $\gamma=13/3$. In reality, however, binary hardening is driven by variety of processes such as interactions with gas and stars, and likely can maintain large eccentricities down to the merger \citep{Fastidio25}. Each of these change the frequency evolution of the binary in a way that is still poorly understood, which directly propagates into large uncertainties of the associated spectral features of the GW background (see, e.g., \citealp{Kelley17, Villalba22}).
These theoretical uncertainties make it difficult to connect the GW spectrum to different astrophysical origins.

The key property that can be used to unambiguously distinguish between binaries and other scenarios is the level of anisotropy. In fact, due to the rapid expansion of the Universe in its earliest evolutionary stages, any GW backgrounds produced at that time are predicted to be indistinguishable from isotropic (see the excellent Section~3 of \citealp{caprini}).
On the other hand, since SMBHs binaries reside in the centres of galaxies, they are likely to follow the distribution of galaxy clusters, producing potentially detectable anisotropy \citep{Ravi12,Cornish14,Cusin25}. Moreover, the detection of continuous waves from an individual SMBH binary with PTAs, which can be first identified as an anisotropic hot spot in the sky, would provide conclusive evidence that SMBH are a significant source of nanohertz GWs. 

The point of this Paper is to promote a framework for GW analysis that treats GW sky maps as fundamental representations of PTA data.
In doing so, we aim for a unified approach to isotropic searches, continuous-wave searches, and measurements of anisotropy.
At the centre of this framework are ``phase-coherent maps'' first proposed in \cite{Gair14,Cornish14}.
These complex-valued maps measure how $+$ and $\times$-polarized strain amplitudes vary with sky location and frequency. We see several benefits to this approach:

\begin{itemize}
    \item \textit{An elegant means of compressing the data.} The sensitivity of pulsar timing increases with time; it is by design a long-term experiment. Current data sets already consist of more than two decades of observations and detailed analyses are computationally expensive. Derived directly from the pulse times of arrival (TOAs), phase-coherent maps preserve the information in the sensitive band of PTAs while excluding less useful high-frequency content. They therefore represent a minimally processed, complete description of the data, which can be losslessly compressed by retaining only the significant low-frequency components.
    
    \item \textit{A single framework for different searches.} In the current paradigm, separate analyses are carried out (1) to characterise the properties of an isotropic background, (2) to search for single resolvable sources, and (3) to search for anisotropy. Compression provided by phase-coherent mapping allows for a somewhat more natural approach, recognizing the fact that GWs are the principal physical quantity and all subsequent analyses seek to characterise it rather than search for unique signals. 
    
    \item \textit{A more transparent treatment of the data.} Historically, concerns about intrinsic pulsar red noise mimicking GW signatures led to the widespread adoption of cross-correlation based likelihoods, where the Hellings--Downs spatial correlations serve as the primary detection statistic. Current mapping methods have followed suit, constructing likelihoods from inter-pulsar correlations (for a comprehensive review see \citealp{Taylor21}). However, as \cite{Haasteren25} has recently shown, likelihoods built from cross-correlations are surprisingly subtle and remain susceptible to systematic errors from noise misspecification. Moreover, auto-power information is implicitly encoded in the cross-correlations, meaning that the separation between ``signal'' and ``noise'' is less clean than it might appear. 
    
\end{itemize}

Phase-coherent mapping sidesteps these issues by working directly with the TOAs in the frequency domain, making the flow of information from data to inference explicit. Noise misspecification must still be addressed, but the framework makes its impact more 
transparent and easier to diagnose.\footnote{The shift from building analyses around cross-correlations to working directly with TOAs is reminiscent of similar transition that took place in the LIGO-Virgo-KAGRA community. Prior to GW150914~\citep{gw150914}, the first gravitational-wave event, the community placed great emphasis  on the quasi-resampling technique of ``time shifts'' to model the background; see, e.g,. \cite{bcr}. The thinking at the time was that this would help guard against false-positive detections. However, once the community became convinced that GW150914 was astrophysical, the emphasis shifted. In order to work out the properties of GW150914 (and subsequent events), researchers employed Gaussian noise models, not the empirical noise distributions calculated with time shifts \citep{gw150914_properties}. We anticipate a similar shift will take place in pulsar timing as the community transitions from establishing the first detection to \textit{characterising} the nature of the background.}

In this work we present a practical implementation of phase-coherent mapping as a universal framework which we call \textsc{MIMOSIS} (coMprehensIve fraMework for mapping the gravitatiOnal wave Sky with pulsar tIming arrayS). 

The remainder of this Paper is organised as follows. 
In Section~\ref{history}, we provide a short history of efforts to map the GW sky and explain the difference between power maps and phase-coherent maps.
In Section~\ref{sec:methods} we outline the methodology. Results are presented and discussed in Section~\ref{sec:results} and Section~\ref{sec:discussion}. A summary is presented in Section~\ref{sec:conclusions}.

\section{A brief history of mapping}\label{history}

Techniques for characterising the angular distribution of the gravitational wave background were first developed in the context of ground-based detector networks. \cite{Allen97} established the foundational technique of correlating pairs of detectors to probe the angular structure of a stochastic background via spherical harmonic decomposition. \cite{Balmer06} introduced the radiometer approach, targeting point-like sources in a pixel basis, while \cite{mitra} adapted maximum-likelihood map-making techniques from cosmic microwave background analysis to deconvolve the detector beam response. \cite{Thrane09} extended this to multiple baselines using singular-value decomposition to regularise the inversion. 
These methods were applied to LIGO data to produce the first (upper-limit) maps of the gravitational-wave sky \citep{s4_radiometer,sph_results}.
All of these methods produced plots of gravitational-wave \textit{power}---proportional to strain squared---as this is the natural estimator of any cross-correlation analysis.

These methods were subsequently adapted for pulsar timing arrays. \cite{Mingarelli13} generalised the isotropic Hellings--Downs overlap reduction function to anisotropic backgrounds in a spherical harmonic basis, and \cite{Taylor13} developed the first 
Bayesian search pipeline for PTA angular structure using this formalism. In both spherical harmonic and pixel-based reconstructions, the recovered power can take unphysical negative values due to noise. This motivated the introduction of square-root spherical harmonic parameterisations by \cite{Taylor20} and \cite{Payne20} for PTAs and ground-based detectors respectively, and by \cite{Banagiri21} for LISA. On the frequentist side, \cite{Pol22} developed a framework with built-in positivity constraints and detection thresholds for anisotropy.

Subsequent work has focused on optimising and understanding the limitations of these methods. \cite{AliHaimoud20} introduced Fisher-matrix-based principal maps to identify the angular modes to which a given array is most sensitive. \cite{Semenzato25}  demonstrated that unresolved small-scale structure biases the recovered angular power spectrum through scale leakage, while \cite{Konstandin25} systematically benchmarked different basis choices and search strategies. \cite{Grunthal25} applied regularised spherical harmonic mapping to the MeerKAT PTA (henceforth MPTA), and later refined this with an adaptive local resolution scheme based on the array's point spread function \citep{Grunthal26}. Two recent developments also include relaxation of the weak-signal limit and per-frequency optimal statistic \cite{Grunthal25, Gersbach25}. Searches for anisotropy have now been carried out by four PTA collaborations \citep{EPTA_anis,NG_anis,Grunthal25,PPTA_anis}, with tentative hotspots reported but no conclusive evidence for anisotropy to date.

All of the these papers are built around the cross-correlation of pulsar pairs.
However, cross-correlation discards phase and polarisation information in the process. An alternative approach---phase-coherent mapping---was proposed by \cite{Gair14} and \cite{Cornish14}, who showed that the full complex polarisation state of the GW sky can be recovered directly from the timing residuals. This preserves strictly more information than power-based methods and provides a unified framework for analysing backgrounds, anisotropy, and individual sources within a single formalism.\footnote{Interestingly, while phase-coherent mapping offers tremendous potential for pulsar timing, it does not offer significant benefits for audio-band observatories. In pulsar timing, the phase measured in a given frequency bin can be related to the parameters of a continuous-wave source. However, audio-band maps are calculated using the average of ${\cal O}(10^5)$ data segments, which has the affect of scrambling phase information. (No such averaging takes place in pulsar timing.)}
The present work is a practical implementation of these ideas.

\section{Methods}\label{sec:methods}
\subsection{Pulsar timing and GWs}

Gravitational waves induce a perturbation to the spacetime metric that can be decomposed into two polarisation modes:
\begin{equation}\label{eq:gw}
h_{ab}(t,\hat{\Omega})
= e^{+}_{ab}(\hat{\Omega})\,h_{+}(t,\hat{\Omega})
+ e^{\times}_{ab}(\hat{\Omega})\,h_{\times}(t,\hat{\Omega}) \,
\end{equation}
where $h_{+}$ and $h_{\times}$ are the strain amplitudes of the $+$ and $\times$ polarisations for a wave
propagating in the $\hat{\Omega}$ direction, and $e^{A}_{ab}(\hat{\Omega})$ are the corresponding
polarisation tensors ($A\in\{+,\times\}$). 

The geometric response of a pulsar to GWs is described with antenna patterns \citep{Sesana10}: 
\begin{equation}\label{eq:antenna}
    \mathcal{F}^{A}(\hat{\Omega}) \equiv \frac{1}{2}\,\frac{\hat{p}^{a}\hat{p}^{b}}{1+\hat{\Omega} \cdot \hat{p}}\,e^{A}_{ab}(\hat{\Omega}) \,,
\end{equation}
where $\hat{p}$ indicate each pulsar's position in the sky. Explicit equations for all the components in Eq.~\ref{eq:gw} and Eq.~\ref{eq:antenna} are presented in Appendix \ref{appA}.

Gravitational waves distort the spacetime at the Earth and at the pulsar and cause a time-dependent Doppler shift (redshift) of the star's pulse frequency $\nu$:\footnote{Here and throughout the paper, repeated indices imply summation unless otherwise stated.}

\begin{equation}
z(t,\hat{\Omega}) \equiv \frac{\nu(t) - \nu_0}{\nu_0}
= \mathcal{F}^A(\hat{\Omega})\Delta h_A(t,\hat{\Omega}) \, ,
\end{equation}
where $\Delta h_A$ denotes the difference between the metric perturbation at the Earth (at time $t_E$ and location $\vec{x}_E$) and at the time and location of the pulsar ($t_P=t_E-L/c$, $\vec{x}_P=\vec{x}_E+L\hat{p}$, where $L$ is the pulsar-Earth distance):
\begin{equation}\label{eq:deltah}
\begin{split}
\Delta h_{A}(t,\hat{\Omega})
=
\int_{-\infty}^{\infty} df\, h_{A}(f,\hat{\Omega})\,
e^{i\Phi_E} \big(1 - e^{-i\Phi_P}\big) .
\end{split}
\end{equation}
Here, $\Phi_E$ and $\Phi_P$ describe the \textit{Earth} and \textit{pulsar} terms, respectively, which can be written as:
\begin{equation}\label{eq:terms}
\begin{split}
\Phi_E &= 2\pi f\left(t-\frac{\hat{\Omega}\cdot\vec{x}}{c}\right), \\
\Phi_P &= 2\pi f L \,\frac{\left(1+\hat{\Omega}\cdot\vec{p}\right)}{c}.
\end{split}
\end{equation}
In deriving Eq.~\ref{eq:deltah}, we assume that the binary signal is approximately monochromatic, so that the GW frequency does not evolve appreciably during the travel time from pulsar to Earth.\footnote{This is typically a reasonable approximation. Consider, for example, a $10^9 M_\odot$ binary observed with Earth-term frequency $\unit[7]{nHz}$ using a pulsar at a distance of $\unit[1]{kpc}$. Depending on the sky location of the binary, the pulsar-term frequency may be higher by as much as $\unit[0.7]{nHz}$, which is less than the typical spectral resolution of pulsar timing arrays $\gtrsim\unit[1]{nHz}$. The approximation begins to break down when we consider very massive binaries, observed at high frequencies, with distant pulsars, and long datasets. The more general version of Eq.~\ref{eq:deltah} is 
\begin{align}
    \Delta h_{ab} = h_{ab}(t_E, \vec{x}_E) - h_{ab}(t_E-\tau, \vec{x}_E) .
\end{align}
Here, $\vec{x}_E$ is the position vector for Earth and $\tau$ is the time delay between the pulsar and Earth.
In this work, we do not attempt to measure the pulsar term.
Instead, we treat it as a source of noise.
Thus, for our purposes, the subtleties of this approximation are not practically important.}

In pulsar timing, the key data product is the set of timing residuals $\delta t_i(t)$ for each pulsar $i$, comprising both the GW signal and noise components $n_i(t)$. In standard timing methods, GWs are treated as a stochastic noise-like process.
However, here we deliberately separate the signal $r$ and noise $n$:
\begin{align}\label{eq:resid_contributions}
    \delta t_i = r_i(t) + n_i(t) .
\end{align}

The timing distortions due to GWs observed by the PTAs in the residuals $r_i(t)$ are related to the redshift integrated over time:
\begin{align}\label{eq:ri}
r_i(t) \equiv & \int_{0}^{t} dt'\, z(t') \nonumber\\
= & \int_{-\infty}^{\infty} df\, \frac{1}{i\,2\pi f} \int_{S^2} d^2
 \Omega \, 
 \mathcal{F}^A(\hat{\Omega})
h_{A}(f,\hat{\Omega})\,
e^{i\Phi_E} \nonumber\\
& 
\big(1- e^{-i\Phi_P} \big) .
\end{align}

All subsequent analysis is carried out in the frequency domain, which offers several practical advantages. First, it is computationally efficient: noise in the timing residuals can be described through its power spectral density and treated with rank-reduced models, so the analysis can be truncated to a relatively small set of Fourier components rather than the full set of time-domain samples. Second, the frequency-domain representation aligns naturally with GW signals (for example, individual SMBH binary systems emit nearly monochromatic waves). Finally, working in frequency space makes phase information explicit and easy to track.
We employ a Fourier design matrix $\bm F$ (defined in Appendix~\ref{appA}) evaluated at discrete observing times in order to transform the residuals to the frequency domain as:
\begin{align}\label{eq:fourier}
    {\delta t} (f_k) = {F}_{km} \, {\delta t}(t_m).
\end{align}

The signal component in the frequency domain can be written as:
\begin{align}\label{eq:rf}
r_i(f) 
= & \frac{1}{i\,2\pi f} \int_{S^2} d^2
 \Omega_{\hat{k}} 
 \mathcal{F}^A(\hat{\Omega})
h_{A}(f,\hat{\Omega})\,
e^{i\Phi_E} .
\end{align}
Since there are two polarisation states ($+,\times$), and since we measure the real and imaginary components of each polarisation state, there are four maps associated with each frequency bin. Each of these components  can be projected onto the sky to reconstruct the signal's angular distribution. This can be achieved by expanding the signal either in a spherical-harmonic basis or with a pixel basis; the choice between the two bases is arbitrary since one can simply transform between bases. In our work we use the pixel basis and divide the sky into equal-area pixels with HEALPix \citep{Gorski05}, which we label with Greek indices $\mu,\nu$. 

Now that our maps are discretised into pixels, we can rewrite Eq.~\ref{eq:rf} in a compact matrix form using Einstein summation notation:
\begin{align}\label{eq:gwresids}
    r_i(f) =  \frac{1}{i\,2\pi f} {\cal F}_{i\nu} \, h_\nu(f) .
\end{align}
where the GW strain map in each frequency is fully characterised by a set of complex Fourier coefficients for each pixel $\nu$:
\begin{align}\label{eq:hnu}
        h_\nu = 
        \begin{pmatrix}
        \text{Re}(h_+(\hat\Omega_{\nu=1})) \\
        \text{Im}(h_+(\hat\Omega_{\nu=1})) \\
        \text{Re}(h_\times(\hat\Omega_{\nu=1})) \\
        \text{Im}(h_\times(\hat\Omega_{\nu=1})) \\
        \vdots
        \end{pmatrix} .
    \end{align} 
To assist with readability, we summarise our notation in Tab.~\ref{tab:notation}.

\begin{deluxetable}{lll}
\tablecaption{Summary of the notation used throughout this work.\label{tab:notation}}
\tablehead{
\colhead{Variable} & \colhead{Meaning} & \colhead{Indices}
}
\startdata
$p_a$   & spatial coordinates & $a,b$ \\
$t_m$   & time & $m,n$ \\
$f_k$   & frequency bin & $k,l$ \\
$r_i$   & pulsar         & $i,j$ \\
$Y_\mu$ & pixel & $\mu,\nu$ \\
$A$     & polarisation state & $+,\times$ \\
\enddata
\end{deluxetable}

\subsection{Noise modelling}
Noise in pulsar timing can be divided into two main classes: white noise (characterized by a flat power spectral density) and red noise (characterized by a power spectral density that rises at low frequencies). The main sources of white noise are radiometer noise, instrumental errors and radio pulse shape variations. Red noise can be separated into chromatic and achromatic components, based on whether or not the noise source depends on radio frequency. The main sources of red noise are pulsar rotation irregularities (achromatic spin noise) and propagation effects from radio waves traversing the ionised interstellar medium---most notably stochastic variations in dispersion measure (chromatic). Additional chromatic contributions can arise from instrumental systematics that vary across observing bands (e.g., band-dependent calibration or pulse profile-evolution misspecification). 

The noise covariance matrix encodes the combined contribution of all noise processes in each pulsar’s residuals and is given by:\footnote{Repeated indices do not imply summation.}
\begin{align}\label{eq:cov}
\mathbf{C}_i \equiv \left\langle \bm{n}_i \bm{n}_i^{\mathsf T} \right\rangle
\end{align}
We transform it to the frequency domain with the Fourier design matrix as:
\begin{equation}
\begin{aligned}
\tilde{\mathbf{C}}_i(f_k) 
&\equiv \left\langle \bm{n}_u^{i\,\dagger}(f_k)\, \bm{n}_v^{i}(f_k) \right\rangle \\
&= \mathbf{F}_{ua}^\dagger \, (\mathbf{C}_i)_{ab}\, \mathbf{F}_{bv}
\end{aligned}
\end{equation}

In practice, the noise covariance matrix for each pulsar $\bm{{C}}_i$ can be written as a sum of white and red noise contributions. Moreover, as we treat noise in each pulsar as independent, the full covariance matrix for all $\rm N_{psr}$ pulsars in the array will be block-diagonal
$\mathbf{C} \equiv \mathrm{diag}\!\left(\bm{{C}}_1, \ldots, \bm{{C}}_{\rm N_{\rm psr}}\right)$. 
We highlight that our covariance matrix is fundamentally different from the ones used in cross-correlation analyses, which include contributions from Earth-term gravitational-wave signals.
There are no Earth-term gravitational-wave signals in our covariance matrix; only noise.

Red noise processes are modelled as Gaussian processes in a Fourier basis. The timing residuals for pulsar $i$ induced by some red noise process can be written as \citep{Taylor21}:\footnote{Repeated indices do not imply summation}
\begin{equation}
    \bm{t}_{\mathrm{RN},i} = \mathbf{F}_i \bm{b}_i
\end{equation}
where $\bm{F_i}$ is again the Fourier design matrix whose columns are sine and cosine functions evaluated at the pulsar's TOAs (defined in Appendix \ref{appA}), 
and $\bm{b}_i$ is the vector of Fourier coefficients of the Gaussian process (i.e., the amplitudes of each sine and cosine component). The Fourier coefficients are drawn from a zero-mean Gaussian prior:
\begin{equation}
    p(\bm{b}_i|\bm{\eta}_i) = \frac{\exp\!\left(-\frac{1}{2}\bm{b}_i^{\mathsf T}
    \mathbf{B}_i^{-1}\bm{b}_i\right)}{\sqrt{\det(2\pi\mathbf{B}_i)}},
\end{equation}
where $\bm{\eta}_i$ denotes the hyper-parameters (e.g. amplitude $\{A_{RN_i}$ and spectral index of the red noise process $\gamma_{RN_i}\}$) and 
$\mathbf{B}_i$ is the prior covariance matrix of the Fourier coefficients. For a 
single pulsar, $\mathbf{B}_i$ is diagonal with entries given by 
the discretised power spectral density:
\begin{equation}
    (B_i)_{kk} = \kappa_i(f_k) 
    = \frac{A_i^2}{12\pi^2}\,\frac{1}{T}
    \left(\frac{f_k}{1\,\mathrm{yr}^{-1}}\right)^{-\gamma_i}\,\mathrm{yr}^{2},
\end{equation}
where $f_k = k/T_{\rm obs}$ are the Fourier sampling frequencies. This straightforwardly extends to multiple red noise processes (e.g., achromatic spin noise 
with spectrum $\kappa_a(f)$ and chromatic DM noise with spectrum $\lambda_a(f)$), in 
which case the diagonal entries of $\mathbf{B}$ are the sum of the individual power 
spectra:
\begin{equation}\label{eq:phi_intrinsic_red}
B_{ak,bj} \;=\; \bigl(\kappa_{ak} + \lambda_{ak} + \cdots\bigr)\,
\delta_{ab}\,\delta_{kj}\,.
\end{equation}
Note that unlike the full prior covariance in standard PTA analyses \citep{Taylor21}, which includes the spatially correlated GWB contribution through the overlap reduction function, our $\mathbf{B}_i$ contains only per-pulsar noise processes. The GW signal enters our framework through the mapping equations, not the 
noise model.

The full noise covariance matrix can therefore be written as
\begin{equation}
\mathbf{C} = \mathbf{N} + \mathbf{T}\,\mathbf{B}\,\mathbf{T}^{\mathsf T},
\end{equation}
where $\bm{N}$ is the block-diagonal white noise covariance matrix 
and $\bm{T} = [\bm{M}\;\bm{F}]$ is the combined design matrix, 
concatenating the timing model design matrix $\bm{M}$ and the 
Fourier design matrix $\bm{F}$ \citep{Taylor21}. The likelihood 
requires the computationally expensive calculation of ${\bm C}^{-1}$, which scales as 
$\mathcal{O}\!\left(N_{\rm psr}^3\,N_{\rm TOA}^3\right)$. However, the dimensionality 
can be significantly reduced by employing the Woodbury matrix identity:
\begin{equation}\label{eq:woodbury}
    (\bm{N} + \bm{TBT}^{\mathsf T})^{-1} = \bm{N}^{-1} - 
    \bm{N}^{-1}\bm{T\Sigma}^{-1}\bm{T}^{\mathsf T}\bm{N}^{-1},
\end{equation}
where $\mathbf{\Sigma} = \mathbf{B}^{-1} + 
\mathbf{T}^{\mathsf T}\mathbf{N}^{-1}\mathbf{T}$.

There are several subtleties to discuss before we move on. Regardless of the method, analysing the GW sky requires \textit{prior} and accurate noise modelling. Maximum likelihood posterior values of the noise parameters are then stored in a covariance matrix in Eq.~\ref{eq:cov} ensuring proper separation of the noise and astrophysical signals. It has been shown that noise misspecification can lead to erroneous inference of the GW properties, even with the use of cross-correlations, and therefore it is of utmost importance that the noise models are carefully validated and well-characterised before any GW analysis is performed \citep{Tiburzi16, Hazboun20,Zic22,DiMarco25}.

That said, for the sake of simplicity, we assume henceforth that the single pulsar noise is well known.
This allows us to focus on making maps.
In subsequent work, we intend to address noise modelling in further detail, as this will be necessary for a production-level pipeline.
While we anticipate that significant development will be required, we also anticipate that it will be possible to leverage existing resources used for cross-correlation and continuous-wave searches.
(Noise modelling is a pre-processing step that is to some extent independent of the subsequent gravitational-wave analysis.)

The pulsar term acts as an additional source of uncorrelated noise if we do not know the distance to the pulsar.\footnote{For an intuitive understanding of how the pulsar term creates noise, we refer the reader to Eq.~\ref{eq:rf}. When the pulsar distance is unknown, the $e^{-i\Phi_P}$ term in that expression adds a random phase to the residuals $r_i(f)$.}
This noise contribution may be accounted for by adding it to the diagonal terms of the covariance matrix \citep{Cornish14}. 
The variance increases accordingly, and the updated covariance matrix becomes
\begin{align}
    C_i \rightarrow C_i +
    {\cal F}_{i \mu}^\dagger \, 
    \sigma_\mu^2 \, \delta t_{\mu\nu} 
    \, {\cal F}_{\nu j} ,
\end{align}
where $\sigma_\mu^2$ is the variance of the GW signal from a given direction. 
In practice, we suspect that this gravitational-wave noise will be largely degenerate with pulsar red noise.
In this work, we do not attempt to model the pulsar term, effectively treating it as noise. We intend to explore the pulsar term in a subsequent publication, which we discuss further in Sec.~\ref{sec:pulsar_term}.

Lastly, clock and solar system ephemeris errors can induce correlated noise between pulsar pairs with monopolar and dipolar patterns. These terms can be calculated separately and added as off-diagonal terms to the covariance matrix. Such correlated noise is still distinguishable from the GWs which are by definition quadrupolar. In this work, for simplicity we assume their contributions are negligible and leave their implementation for the second release. 

\subsection{Mapping}
\subsubsection{Likelihood}
The likelihood of the data in a single frequency bin is
\begin{align}\label{eq:lh}
    {\cal L}(\delta t_i | h_\nu) = \frac{1}{ \sqrt{\text{det}(2\pi C)}}
    e^{-\frac{1}{2}
    (\delta t_i - \mathcal{F}_{i\nu}h_\nu)^\dagger
    C_{ij}^{-1}
    (\delta t_j - \mathcal{F}_{j\mu}h_\mu)
    } .
\end{align}
Here, $\delta t_i$ is the Fourier transform of the times of arrival $\delta t (f_i)$ given in Eq.~\ref{eq:fourier}.
Our goal is to find the gravitational-wave sky map $h_\nu$, which maximises the likelihood.
The solution may be obtained analytically.
Following \cite{Thrane09,mitra}, we write the maximum-likelihood solution in terms of a ``dirty map'' and a Fisher matrix.

\subsubsection{Fisher matrix and dirty map}
The dirty map, denoted $X_\nu$, describes the gravitational-wave sky convolved with the detector response:\footnote{We stated above that phase-coherent maps are ``lossless.'' We can see this mathematically in Eq.~\ref{eq:X}. If we decompose the sky map such that the number of basis elements is equal to the number of pulsars, then ${\cal F}_{i\nu}$ is square. In this case Eq.~\ref{eq:X} is \textit{invertible}, and so the frequency-domain data can be entirely reproduced from the dirty map. The same basic idea holds when the number of sky-map pixels exceeds the number of pulsars.}
\begin{align}\label{eq:X}
    X_\nu = \mathcal{F}_{i\nu}^\dagger C_{ij}^{-1} \delta t_j ,
\end{align}
The Fisher matrix $M_{\mu\nu}$ describes the covariance of strain measured in different parts of the sky (and in different polarisation states):
\begin{align}
    M_{\mu\nu} = \mathcal{F}_{i\mu}^\dagger C_{ij}^{-1} \mathcal{F}_{j\nu} .
\end{align}
It is the covariance matrix for the dirty map so that 
\begin{align}
    M_{\mu\nu} = \langle X_\mu X_\nu \rangle - \langle X_\mu \rangle \langle X_\nu \rangle .
\end{align}

\subsubsection{Clean map}
The dirty map and Fisher matrix may be combined to produce a \textit{clean} map, which undoes the detector response to yield estimates of the gravitational-wave strain in different patches of sky:
\begin{equation}\label{eq:clean_map}
    P_\mu = M_{\mu\nu}^{-1} X_\nu .
\end{equation}
This is the maximum-likelihood solution for Eq.~\ref{eq:lh}.
It is an estimator for $h_\nu$ in Eq.~\ref{eq:hnu}.
The clean-map uncertainty is characterised by a covariance matrix given by the inverse Fisher matrix:
\begin{equation}
    M_{\mu\nu}^{-1} = \langle P_\mu P_\nu \rangle - 
    \langle P_\mu \rangle
    \langle P_\nu \rangle .
\end{equation}

A practical difficulty arises when inverting the Fisher matrix, as it is 
rank-deficient: an array of $N_{\mathrm{psr}}$ pulsars provides at most 
$N_{\mathrm{psr}}$ independent constraints per frequency bin, while the 
Fisher matrix describes $2N_{\mathrm{pix}}$ sky parameters (two 
polarizations at each pixel). For current PTAs, 
$N_{\mathrm{psr}} \ll 2N_{\mathrm{pix}}$, and the Fisher matrix must be 
regularised, for instance, via truncated singular-value decomposition, retaining only the most well-constrained eigenmodes.
When the Fisher matrix is ill-conditioned there are some modes (patterns on the sky) that the PTA is relatively insensitive to.
If we naively invert the Fisher matrix, the resulting clean map is dominated by the uncertainty in these unresolvable modes.
Following from \cite{mitra,Thrane09}, we compute a regularised Fisher matrix $\widetilde{\bm{M}}^{-1}$ using a truncated pseudoinverse via singular value decomposition. 

By using a regularized inverse, we discard the poorly-measured modes in order to produce sky maps of the resolvable modes.
With the poorly-resolved modes gone, we are able to reconstruct features of the gravitational-wave sky.
There is a cost, however; the regularised maps are \textit{biased} because they do not include the poorly-resolved modes. We discuss this in detail while presenting our results in Sec.~\ref{sec:sky_maps}.

In this work, we determine the number of significant eigenmodes similarly as in \cite{Grunthal25}, i.e., by calculating the eigenspectrum and visually inspecting the recovered maps, and leave more sophisticated treatment for future work (see, e.g., \citealp{Grunthal26}). We adopt a fixed regularisation retaining $30\%$ of the eigenmodes. The optimal truncation is direction-dependent---at poorly-covered locations, retaining a modestly higher fraction (e.g., $50\%$) can improve the clean map S/N, but the gain is marginal, and so we use a fixed value to 
ensure consistency.

\subsubsection{Radiometer map}
In addition to the clean map solution, it is also useful to plot \textit{radiometer maps}.
The radiometer map is constructed to \textit{ignore} the covariance between pixels. 
The difference between the radiometer and clean maps is that the former is well suited for a point estimate of the strain coming from pixel $\mu$ assuming that there is no other GW signal on the sky.
However, it does not provide reliable reconstructions of extended sources.
Since it does not take into account covariance between pixels, there is no need to regularize the Fisher matrix.
We shall see that it is useful to make both clean maps and radiometer maps in order to understand the distribution of gravitational waves on the sky.

In order to obtain the radiometer map, only diagonal elements of \textbf{M} are inverted, which yields:\footnote{The repeated indices in radiometer map equations do not imply summation.}
\begin{equation}
    \eta_\mu = X_\mu / M_{\mu\mu} .
\end{equation}
The uncertainty associated with $\eta_\mu$ is given by:
\begin{equation}\label{eq:fisher_inverse}
    \sigma_\mu = (M_{\mu\mu})^{-1/2} .
\end{equation}

\subsubsection{Signal-to-noise ratio}\label{sec:snr}
The clean map and the radiometer map can both be used to calculate signal-to-noise ratio (S/N) maps.
For the radiometer, the S/N is given by
\begin{align}
    {\rm S/N}_{\mu}^{\eta} = \eta_\mu/\sigma_{\mu}^{\eta} .
\end{align}
For the clean map, the S/N map is given by
\begin{align}
    {\rm S/N}_{\mu}^{P} = P_\mu/\sigma_{\mu}^{P} .
\end{align}
These S/N map can have both positive and negative values, depending on the phase of the GW signal coming from a given pixel and noise. 
Noise fluctuations tend to produce S/N values with $\pm 3$ of zero, though the exact distribution of S/N due to noise depends on the details of the PTA.
Large departures from S/N=0 are associated with GWs, though care is required to estimate the statistical significance (see \citealp{sph_results,Grunthal25}) 

\subsubsection{Summary}
\begin{itemize}
    \item The dirty map $X_\mu$ is the primary object from which subsequent quantities are constructed.
    One can regard it as a raw representation of the data.
    
    \item The radiometer map $\eta_\mu$ represents our best guess about the strain coming from different directions in the sky --- \textit{if we assume the signal comes from only one direction}. 
    
    \item The clean map $P_\mu$ represents our best guess for the strain coming from the entire sky, taking into account the fact there may be multiple (extended) sources. 
\end{itemize}

\section{Results}\label{sec:results}
\subsection{Simulation setup}
We test the performance and capabilities of our framework with two sets of simulations which are summarised in Table~\ref{tab:sims}. The first one is similar to the current IPTA data set with 10 years of total observing baseline (the real IPTA baseline is around 20 years, but we use a shorter one for simplicity), and 100 pulsars with TOA errors in the range of $\unit[0.1-20]{\mu s}$ (with median values of $\unit[0.5-5]{\mu s}$ per pulsar). 
The distribution of pulsars in the sky is isotropic.

The second simulation set is based on the 83 pulsars observed by the MPTA. In this case, pulsar positions, timing models, TOA uncertainties and noise posteriors from single pulsar analyses are taken from the 4.5 year data release \citep{Miles25} and used to generate a simulated array with the total observing time of 6 years to mimic the upcoming MPTA data release. Individual TOA errors are sometimes much higher than for the IPTA simulation (tens of $\mu$s), however the  median TOA error is no larger than $\unit[5-6]{\mu s}$ across all pulsars. 

For the sake of simplicity, in both sets of simulations, we assume the timing model is perfect. We then inject white and red noise (randomly sampled from appropriate distributions) along with an isotropic GW background and/or individual continuous GW sources. Parameters of the injections are summarised in Table~\ref{tab:injections}. In the case of the MPTA simulation, noise is sampled from posteriors published for the latest data release. We use \texttt{libstempo} \citep{Vallisneri20} to clean the residuals and inject the signals.

\begin{deluxetable}{llll}
\tablecaption{Summary of the simulations in terms of number of pulsars $\rm N_{psr}$, median TOA errors per pulsar $\sigma_{\rm TOA_i}$ and total observing baseline $\rm T_{obs}$. 
\label{tab:sims}}
\tablehead{
\colhead{Name} & \colhead{$N_\text{psr}$} & \colhead{$\mathrm{median}(\sigma_{\rm TOA_i}) [\mu$s]} & \colhead{$T_\text{obs}$ [yr]}
}
\startdata
IPTA        & 100 & $\sim$~0.5--5  & 10 \\
MPTA        & 83  & $\sim$~5     & 6 \\
\enddata
\end{deluxetable}

When injecting the continuous GW source, we set its GW frequency $f_{\rm CGW}$ to match the second Fourier-frequency bin of each PTA simulation. This corresponds to $\sim\unit[6]{nHz}$ for the IPTA case and $\sim\unit[10]{nHz}$ for the MPTA case. 
We adopt an inclination angle of $\iota = \pi/2$ (edge-on), a polarization angle of $\psi = 0$, and an initial GW phase of $\phi_0 = \pi/2$.
This ensures that the complex signal is entirely real and that the polarization is entirely $+$, which in turn makes it straightforward to assess leakage from real to imaginary and from $+$ to $\times$.

Throughout the main analysis, we adopt the Earth-term approximation, in which the pulsar term contribution to the timing residuals is not modelled in the mapping framework. All results presented in the proceeding sections are based on simulations where the pulsar term is also excluded from the signal injection, ensuring a consistent test of the Earth-term mapping pipeline. To assess the impact of this approximation on realistic signals that do contain a pulsar term, we inject continuous GW signals including the pulsar term and examine the recovered maps and present the results in Appendix~\ref{appB}.

\begin{deluxetable}{ll}
\tablecaption{Injection parameters. In the case of the MPTA data set, intrinsic pulsar noise parameters are based on the single pulsar noise analysis from the recent data release, whereas for the IPTA simulation, noise was sampled from appropriate distributions as listed in this table.
\label{tab:injections}}
\tablehead{
\colhead{Parameter} & \colhead{Distribution}}
\startdata
EFAC                      & uniform(0.8, 1.2)    \\
$\rm log_{10}EQUAD$       & uniform(-8.5, -5)      \\
$\rm log_{10}A_{\rm RN}$  & uniform(-20, -12) \\
$\rm \alpha_{\rm RN}$     & uniform(1, 6) \\\\
$\rm A_{ GWB}$            & $2.0\times10^{-15}$ \\
$\rm \alpha_{ GWB}$       & 2/3 \\ \\
$\log_{10}(M_{\rm chirp}/M_\odot)$   & 9 \\
$\rm D_{L}$               & 15~Mpc \\
$\rm f_{CGW}$             & $\unit[6,10]{nHz}$ \\
$\iota$                   & $\pi/2$ \\
$\psi_0$                  & $0$ \\
$\phi_0$                  & $\pi/2$ 
\enddata
\tablecomments{The variables EFAC and EQUAD are white noise parameters that rescale the TOA uncertainties via $\sigma_a'=\rm EFAC^2(\sigma_a^{2}+{\rm EQUAD}^{2})$. The intrinsic red noise power-law is parameterised by amplitude $A_{\rm RN}$ (referenced at $f_{\rm ref}=1~{\rm yr}^{-1}$) and spectral index $\alpha_{\rm RN}$. The stochastic GW background is described by amplitude $A_{\rm GWB}$ (also at $1~{\rm yr}^{-1}$) and spectral index $\alpha_{\rm GWB}$. For the continuous GW injection, $M_{\rm chirp}$ is the SMBH binary chirp mass, $D_L$ the luminosity distance, $f_{\rm cgw}$ the GW frequency, and $\iota$ is the orbital inclination angle.}
\end{deluxetable}

Our HEALPixels are parametrised by the variable \texttt{nside}.
The number of pixels is given by $N_{\rm pix} = 12\times \texttt{nside}^2$.
The solid angle of each pixel is given by $4\pi/N_{\rm pix}$.
For both of our simulations we choose ${\tt nside}=4$ corresponding to 192 pixels, which ensures that we have approximately the correct resolution for our network of $\sim$100 pulsars.\footnote{One can choose to over-resolve the sky with an arbitrarily large number of pixels. We suggest, however, that it is best to approximately match the resolution to the number of pulsars so that the pixel size provides some visual cue for the intrinsic resolution; see \cite{Grunthal26} for a more nuanced discussion of the choice of pixel size.}  

We upscale the \texttt{HEALPix} map plots by transforming the original pixelised map to spherical-harmonic coefficients $a_{\ell m}$ up to the source resolution limit
($\ell_{\max}=3, {\tt nside}_{\rm src}-1$). Then we synthesise a new map on a finer \texttt{HEALPix} grid (${\tt nside}_{\rm plot}=16$) from the same band-limited $a_{\ell m}$.This produces a smoother-looking map, but it does not add new angular information beyond the original $\ell_{\max}$.

Our implementation of the phase-coherent mapping produces several types of sky maps, which we present in the following section. The primary data products are four maps corresponding to the real and imaginary parts of each polarisation: $\mathrm{Re}(h_+)$, $\mathrm{Im}(h_+)$, $\mathrm{Re}(h_\times)$, $\mathrm{Im}(h_\times)$. These can be displayed in units of dimensionless strain (Fourier-domain amplitude) or as S/N as defined in Section~\ref{sec:snr}. 
However, from these we also derive total power maps:
\begin{align}
    P = |h_+|^2 + |h_\times|^2 ,
\end{align}
which can likewise be displayed in terms of Fourier domain strain amplitude ($A = \sqrt{P}$) or S/N.
While this paper prioritizes the creation of maps with both phase and polarization content, we include below a number of power maps as well.
The power maps are useful because---by combining the information in four separate phase-coherent maps---they yield more obvious point sources, which is convenient for illustrative purposes.
They are also useful for comparing with previous observational results, e.g., \cite{Grunthal25}, which is all based on power maps.

For maps containing a continuous GW source, we mark the sky location of the injection with a black hexagon along with the most statistically significant patch of sky (indicated with a circle with a white cross in its centre).
In order to calculate the most significant patch of sky, we determine the set of pixels adjacent to the peak-S/N pixel that have S/N within one of the maximum value on the total power map.\footnote{This metric to determine the most significant patch of sky is intended merely to guide the eye, showing where the signal is loudest. It is not indicative of the width of the point spread function.}

\subsection{Sky maps}\label{sec:sky_maps}
\begin{figure*}
  \centering
  \begin{subfigure}[b]{0.48\textwidth}
    \centering
    \includegraphics[width=\textwidth]{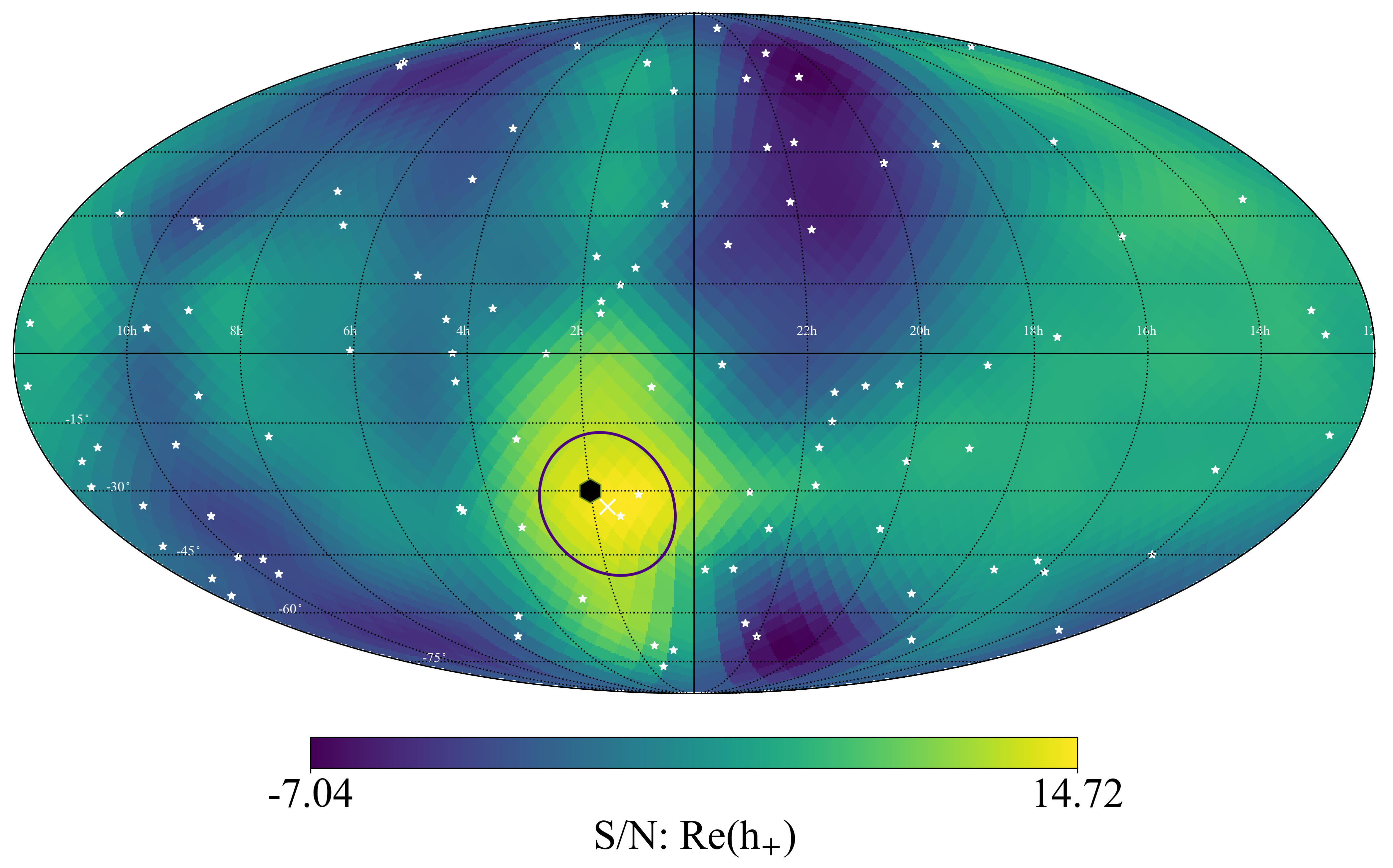}
  \end{subfigure}
  \hfill
  \begin{subfigure}[b]{0.48\textwidth}
    \centering
    \includegraphics[width=\textwidth]{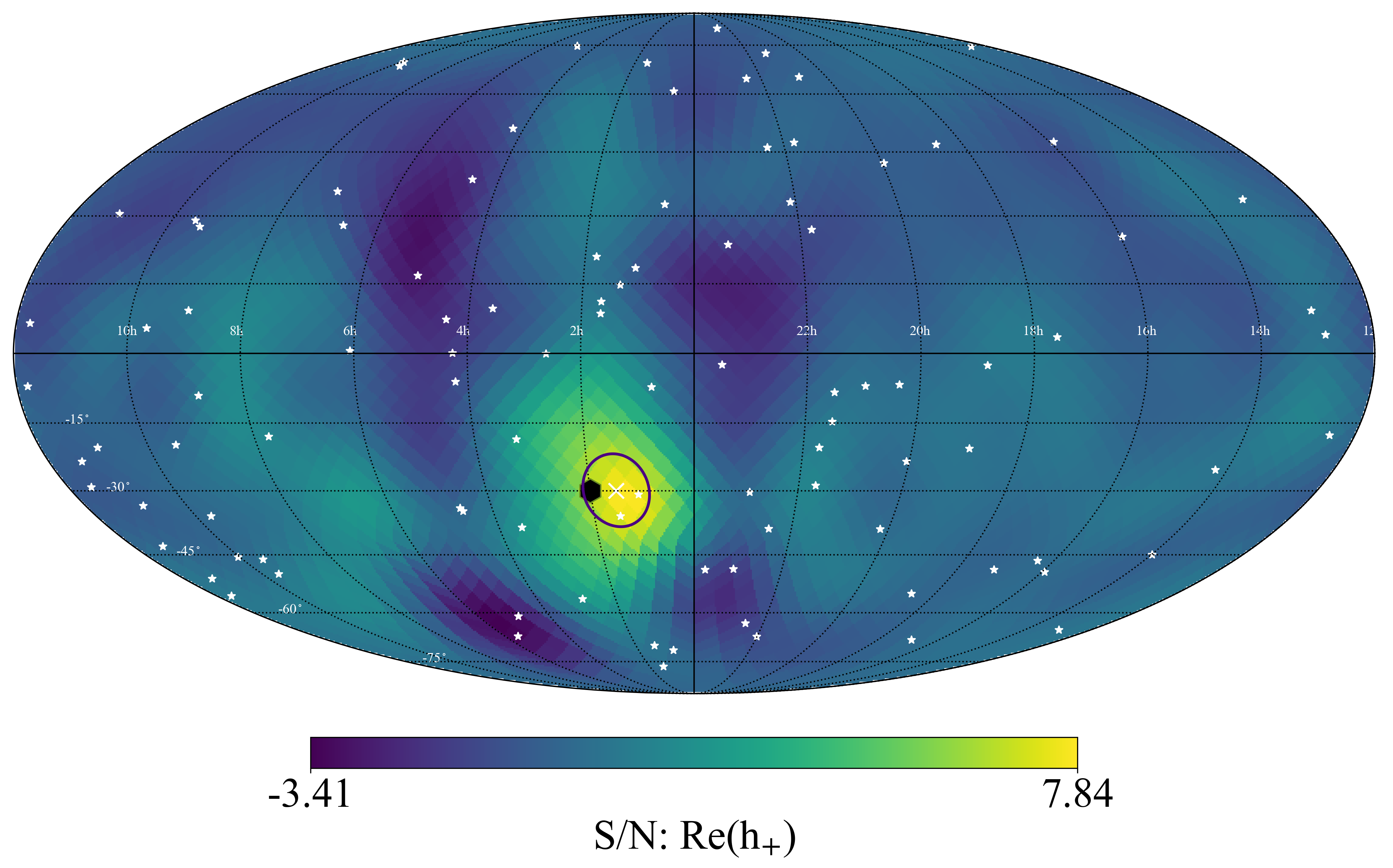}
  \end{subfigure}
  \vspace{0.5em}
  \begin{subfigure}[b]{0.48\textwidth}
    \centering
    \includegraphics[width=\textwidth]{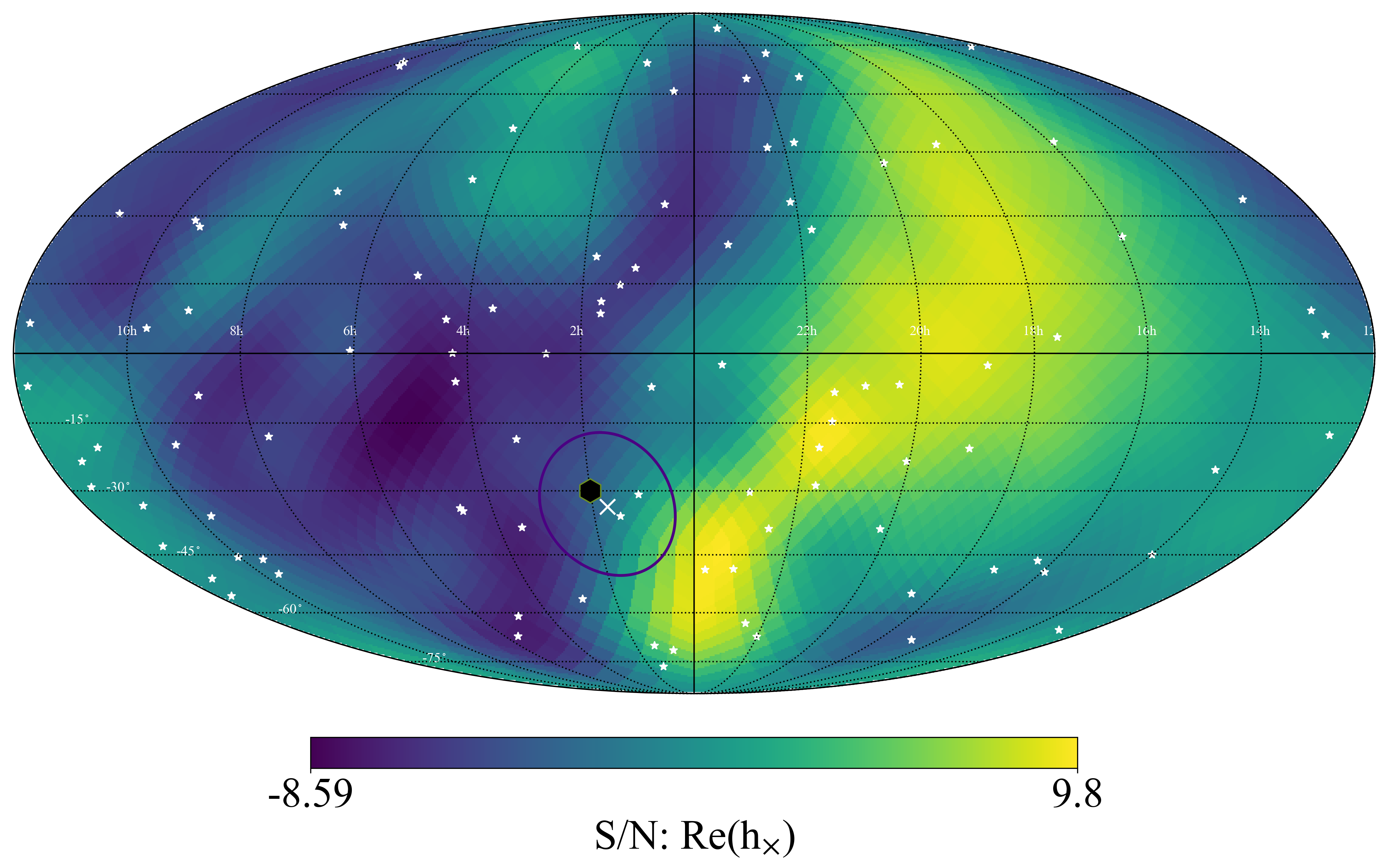}
  \end{subfigure}
  \hfill
  \begin{subfigure}[b]{0.48\textwidth}
    \centering
    \includegraphics[width=\textwidth]{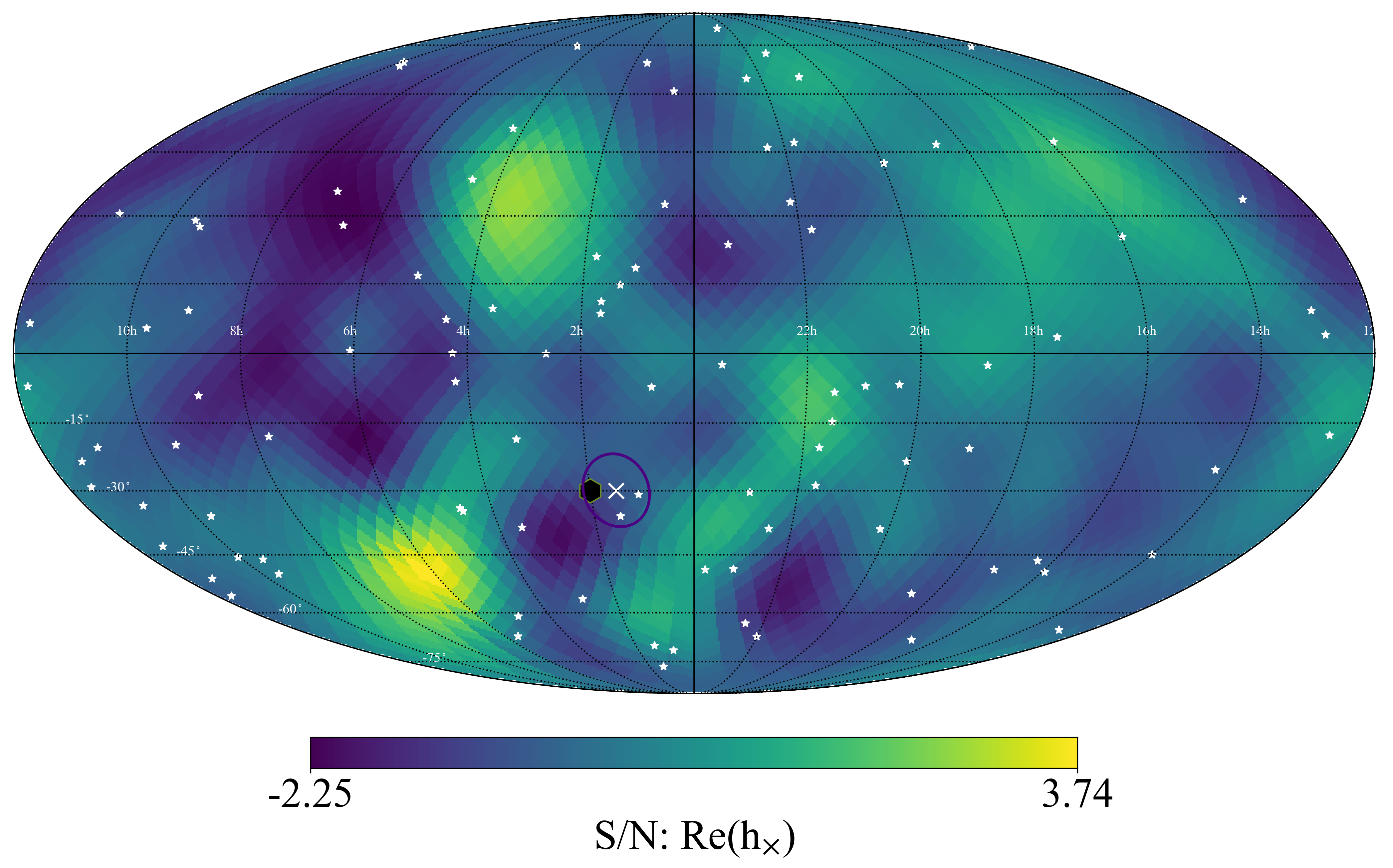}
  \end{subfigure}
  \caption{Radiometer (left) and clean (right) maps of the real components of two polarisation states $h_+$ (top) and $h_{\times}$ (bottom). The injected continuous GW source is defined to emit entirely in $Re(h_+)$ and any strong signal in other channels constitutes leakage (bottom-left). The black hexagon indicates the injected source position, whereas the dark circle shows the recovered high S/N area.}
  \label{fig:hplus_leakage}
\end{figure*}

Figure~\ref{fig:hplus_leakage} shows the S/N maps in two polarisation channels---$\mathrm{Re}(h_+)$ and $\mathrm{Re}(h_\times)$ in the top and bottom rows, respectively---for a single continuous GW source injected at the bottom-left sky location using the IPTA-like array. Since only $\mathrm{Re}(h_+)$ was injected, any signal appearing in the other channels constitutes leakage. We compare radiometer (left column) and clean (right column) maps.

In the radiometer maps, significant sky-wide leakage is visible in the cross polarisation, arising from the geometric correlation between the $F_+$ and $F_\times$ antenna patterns---the sidelobes of one polarisation channel project onto the other at sky locations away from the source. The imaginary components of both polarisations remain consistent with noise, confirming that the spectral separation between the real and imaginary Fourier quadratures is minimal, while the separation between h$_+$ and h$_\times$ is not. Importantly, at the source pixel itself, the leakage is much smaller than the sky-wide maximum, as the antenna patterns are approximately orthogonal for the same sky direction. This means that even in the absence of a clear hot spot, the radiometer strain amplitude at the true source pixel remains a meaningful measurement of the injected signal.
 
The clean map (right column) substantially reduces the polarisation leakage by inverting the full Fisher matrix, which accounts for the off-diagonal correlations between the h$_+$ and h$_\times$ states. In the clean maps, the signal mapped in the cross polarisation is consistent with noise across the entire sky.

\begin{figure}
  \centering
  \begin{subfigure}[b]{0.48\textwidth}
    \centering
    \includegraphics[width=\textwidth]{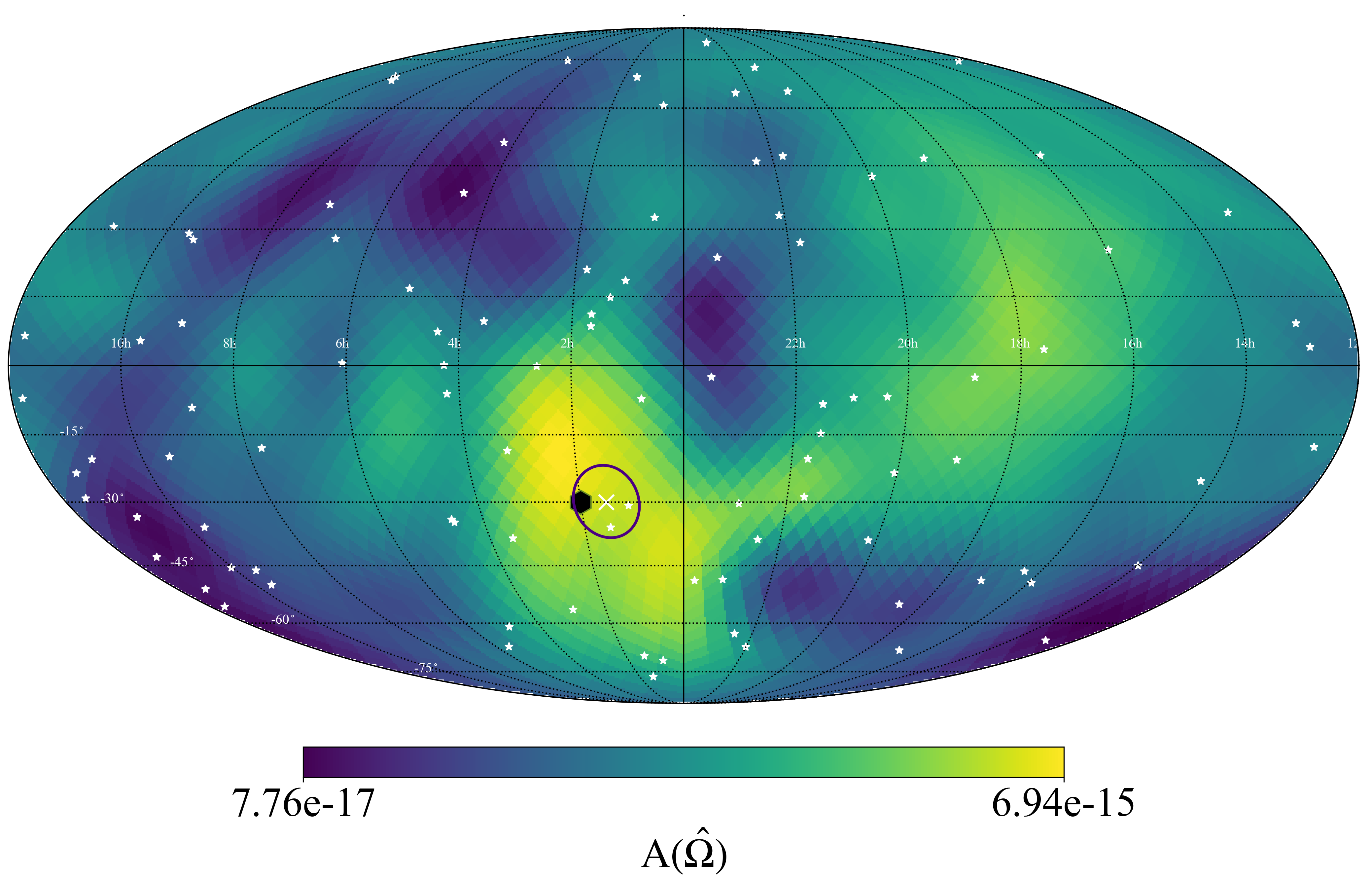}
    \caption{Square root of the radiometer power map in units of dimensionless strain.}
    \label{fig:left}
  \end{subfigure}
  \hfill
  \begin{subfigure}[b]{0.48\textwidth}
    \centering
    \includegraphics[width=\textwidth]{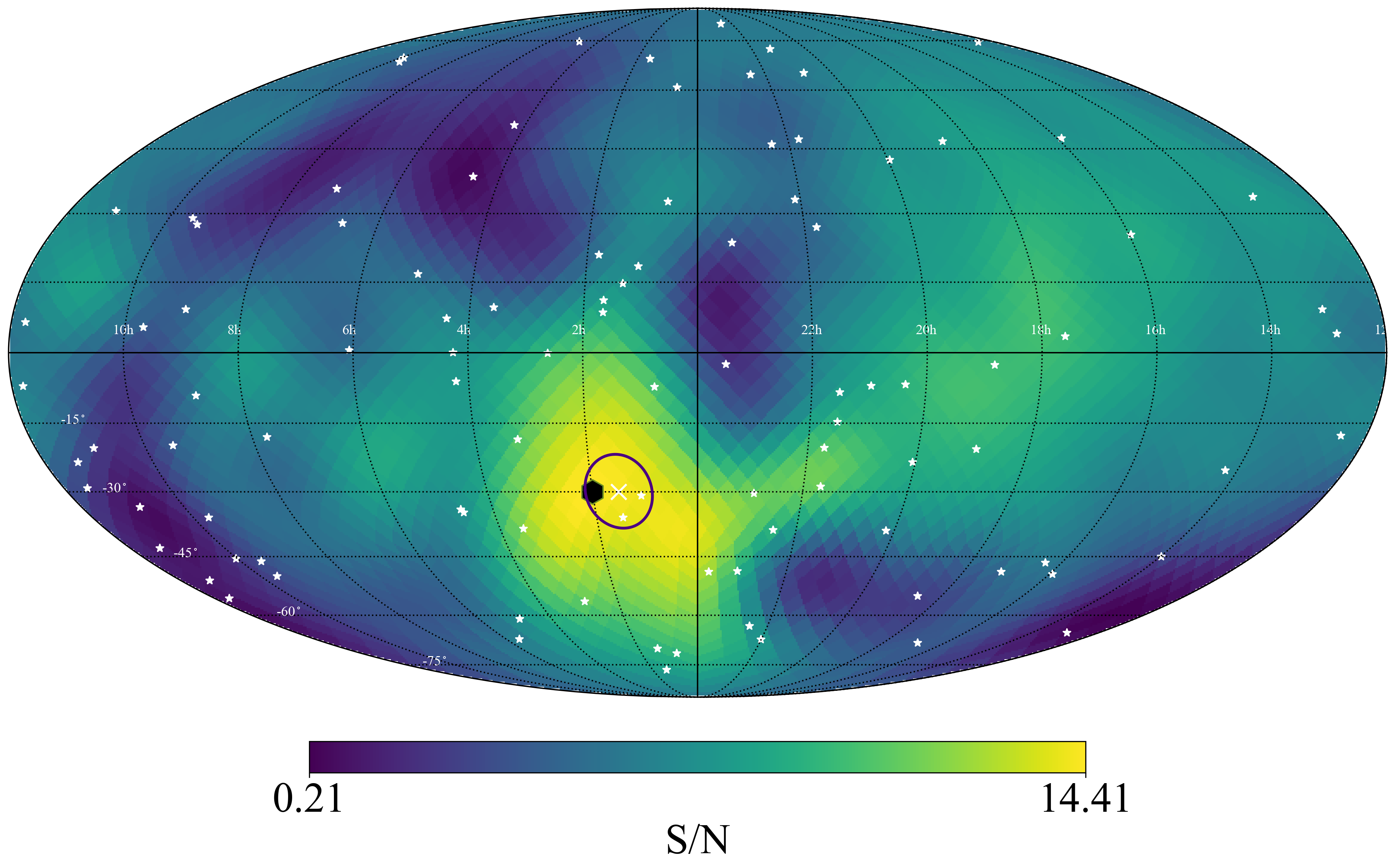}
    \caption{Radiometer S/N.}
    \label{fig:right}
  \end{subfigure}
  \label{fig:both}
  \hfill
  \begin{subfigure}[b]{0.48\textwidth}
    \centering
    \includegraphics[width=\textwidth]{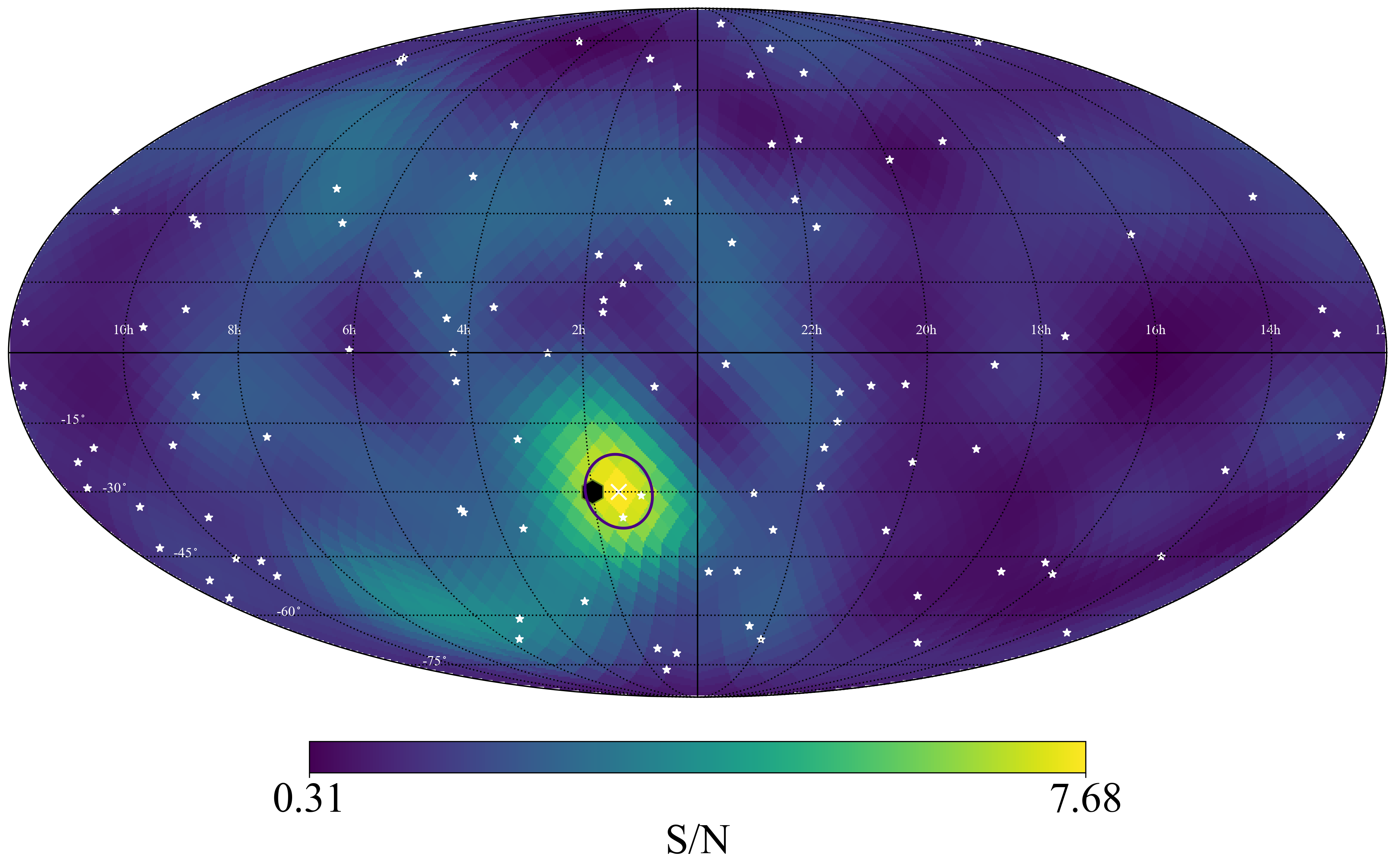}
    \caption{Clean S/N.}
    \label{fig:right}
  \end{subfigure}
  \caption{Total power maps for the IPTA simulation with a continuous GW signal. The dark circle in each plot corresponds to highest S/N region (in this case single pixel) identified in the clean total power map. The black hexagon indicated the true position of the injected GW source.}
  \label{fig:total_power}
\end{figure}

Figure~\ref{fig:total_power} shows the total power maps for the same simulation. Three types of maps are presented: square root of the radiometer total power map, the radiometer S/N map, and the clean S/N map. The most statistically significant patch of sky, determined from the clean map, as described in the previous Section, is indicated with a circle. 
To within a few percent, the radiometer map recovers the correct amplitude $A$ at the source pixel matching the injected sky location. However, the point source produces structure throughout the sky since the radiometer map is not designed to provide an accurate sky map. In contrast, the clean map correctly identifies the source pixel as the sky-wide S/N maximum. However, the clean map S/N is lower than that of the radiometer because the clean map simultaneously estimates a large number of parameters, diminishing the constraining power on any individual pixel. 
This is reflected in the drop in peak S/N between the radiometer and clean maps. 

\begin{figure}
  \centering
  \begin{subfigure}[b]{0.48\textwidth}
    \centering
    \includegraphics[width=\textwidth]{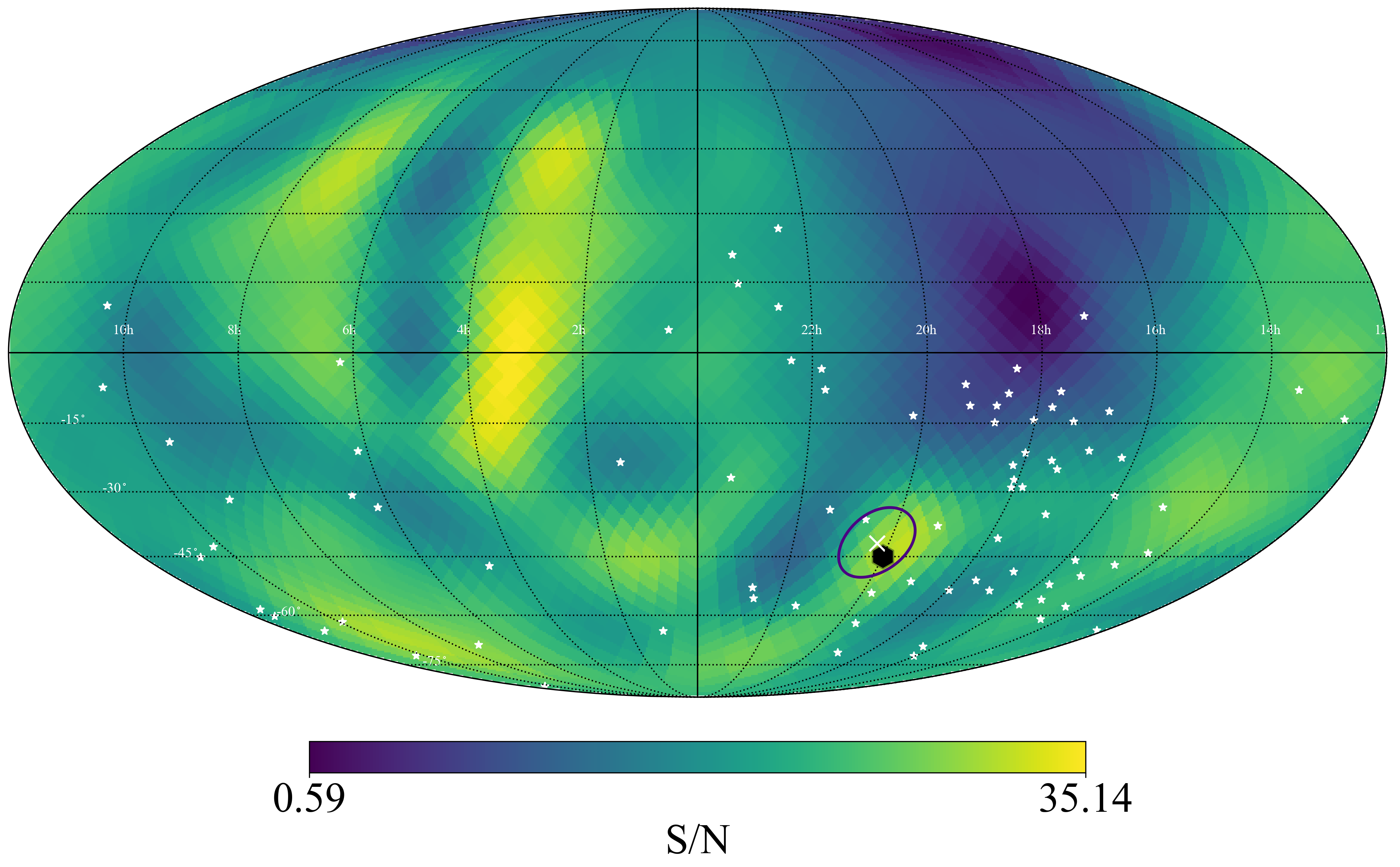}
    \caption{Radiometer S/N.}
  \end{subfigure}
  \hfill
  \begin{subfigure}[b]{0.48\textwidth}
    \centering
    \includegraphics[width=\textwidth]{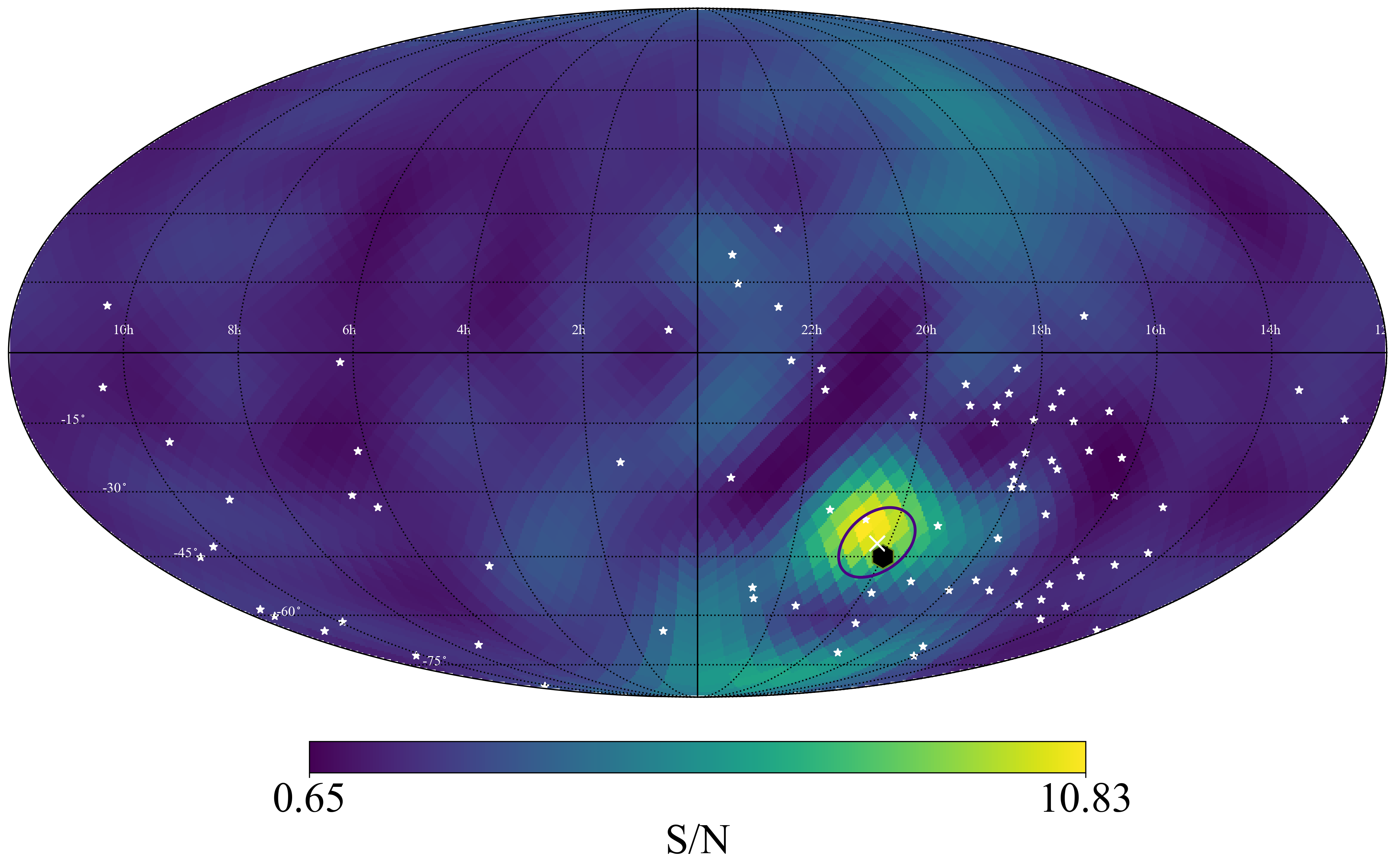}
    \caption{Clean S/N.}
  \end{subfigure}
  \caption{Total power radiometer and clean maps for the MPTA simulation, demonstrating the need for regularisation when localising the source. The dark circle in both maps shows the highest S/N region (here it is again a single pixel) with true source position denoted with a black hexagon.}
  \label{fig:MK_radio_clean}
\end{figure}

Figure~\ref{fig:MK_radio_clean} highlights the complementary roles of the radiometer and clean maps for a more realistic PTA configuration. It shows the recovery of a continuous GW source at the best-covered sky location (bottom-right) using the MPTA-like array, which features a non-isotropic pulsar distribution in the sky and more heterogeneous TOA uncertainties than in the IPTA simulation. Two total-power S/N maps are shown: the radiometer and the clean map. The radiometer map achieves high peak S/N ($\sim 35$) but the signal is spread across the sky making unambiguous source localisation impossible from the radiometer alone. The clean map, by contrast, produces a compact hot spot at the correct source location.

This example illustrates a key result: despite the visually uninformative radiometer map, the strain recovered at the correctly localised pixel is robust. At the best-covered location, the $\mathrm{Re}(h_+)$ recovery fraction is ${\sim}\,97\%$. The clean map provides the localisation, the radiometer provides the amplitude---together they yield both.

\begin{figure*}
  \centering
  \begin{subfigure}[b]{0.48\textwidth}
    \centering
    \includegraphics[width=\textwidth]{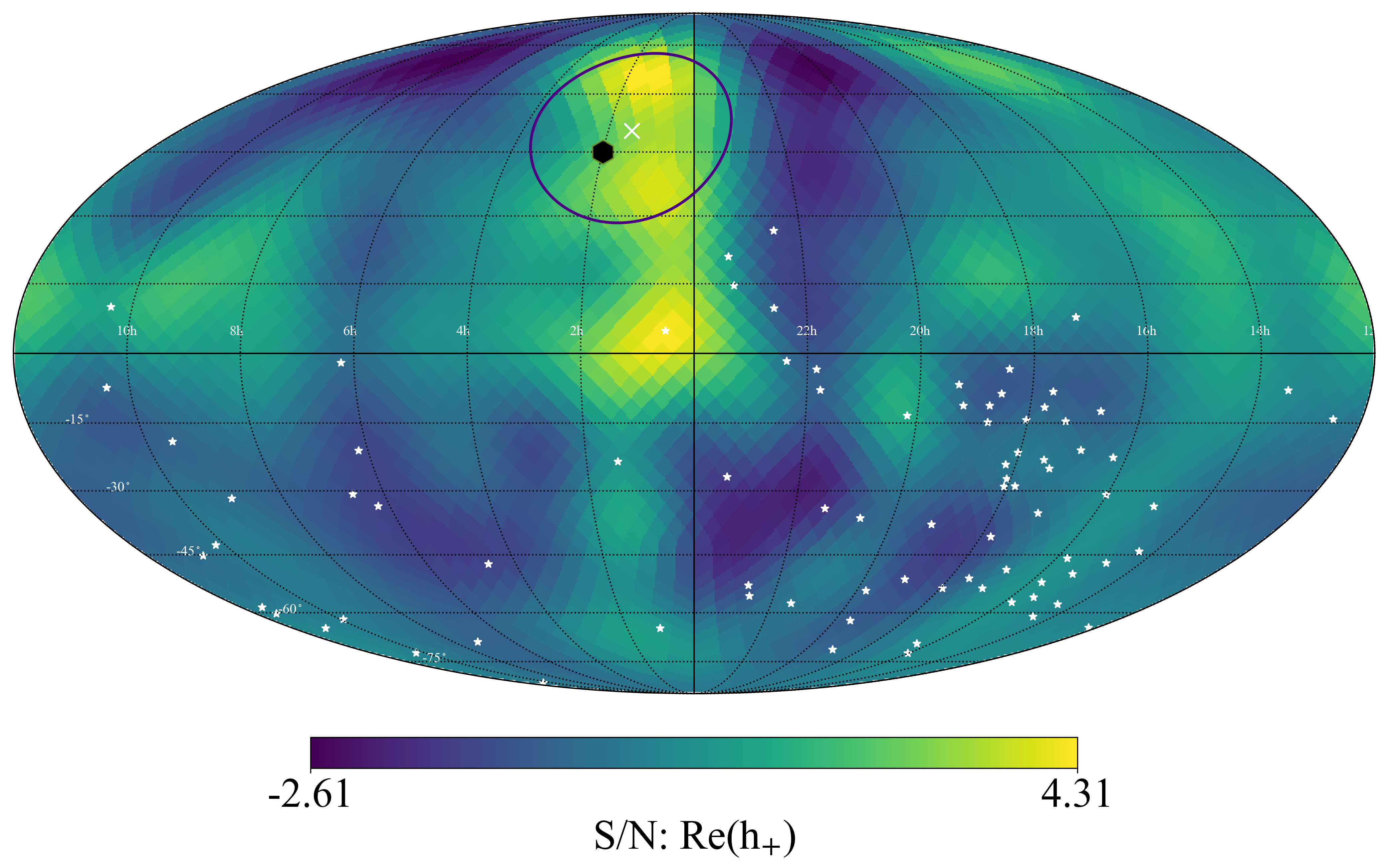}
    \label{fig:tl}
  \end{subfigure}
  \hfill
  \begin{subfigure}[b]{0.48\textwidth}
    \centering
    \includegraphics[width=\textwidth]{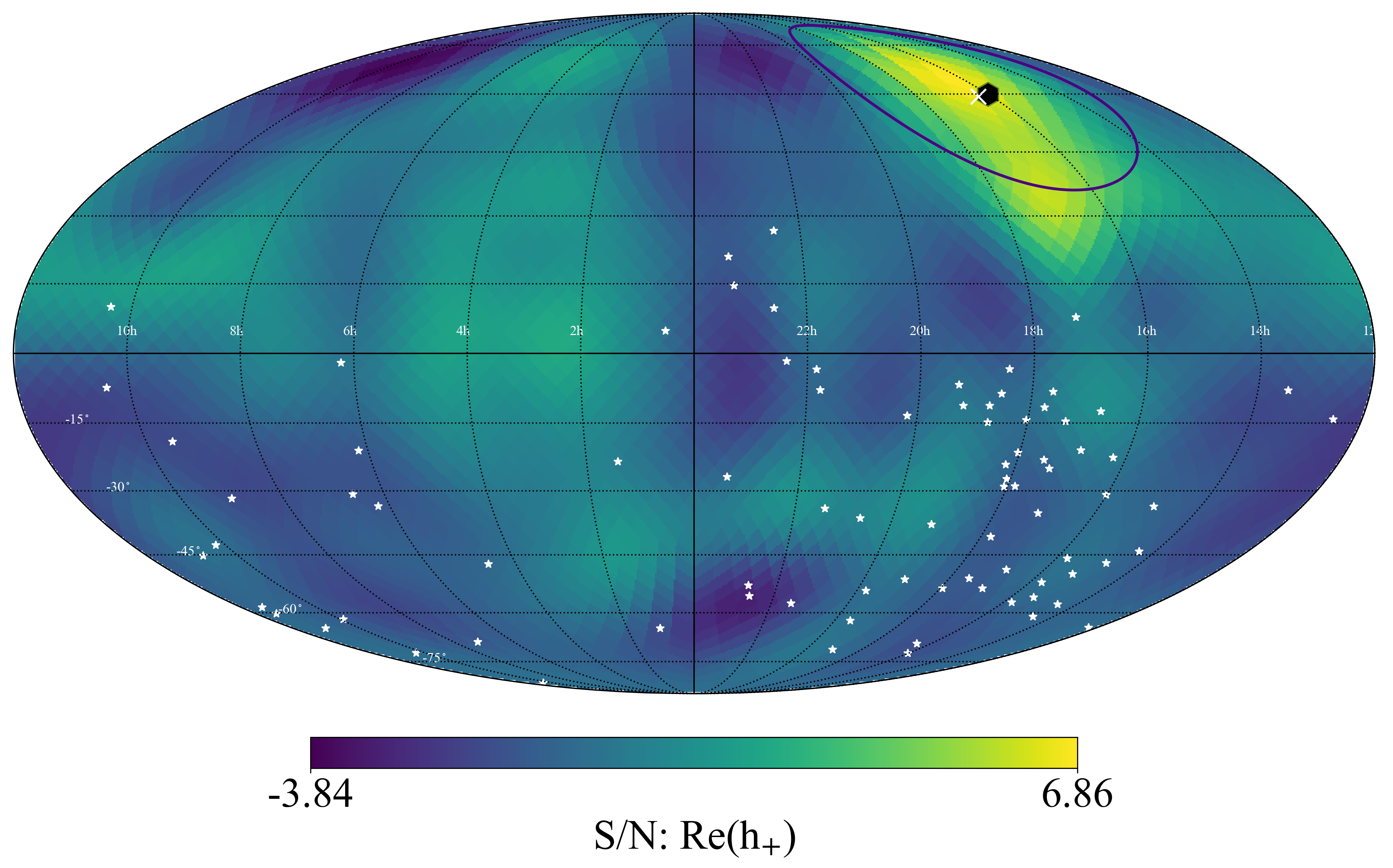}
  \end{subfigure}
  \vspace{0.5em}
  \begin{subfigure}[b]{0.48\textwidth}
    \centering
    \includegraphics[width=\textwidth]{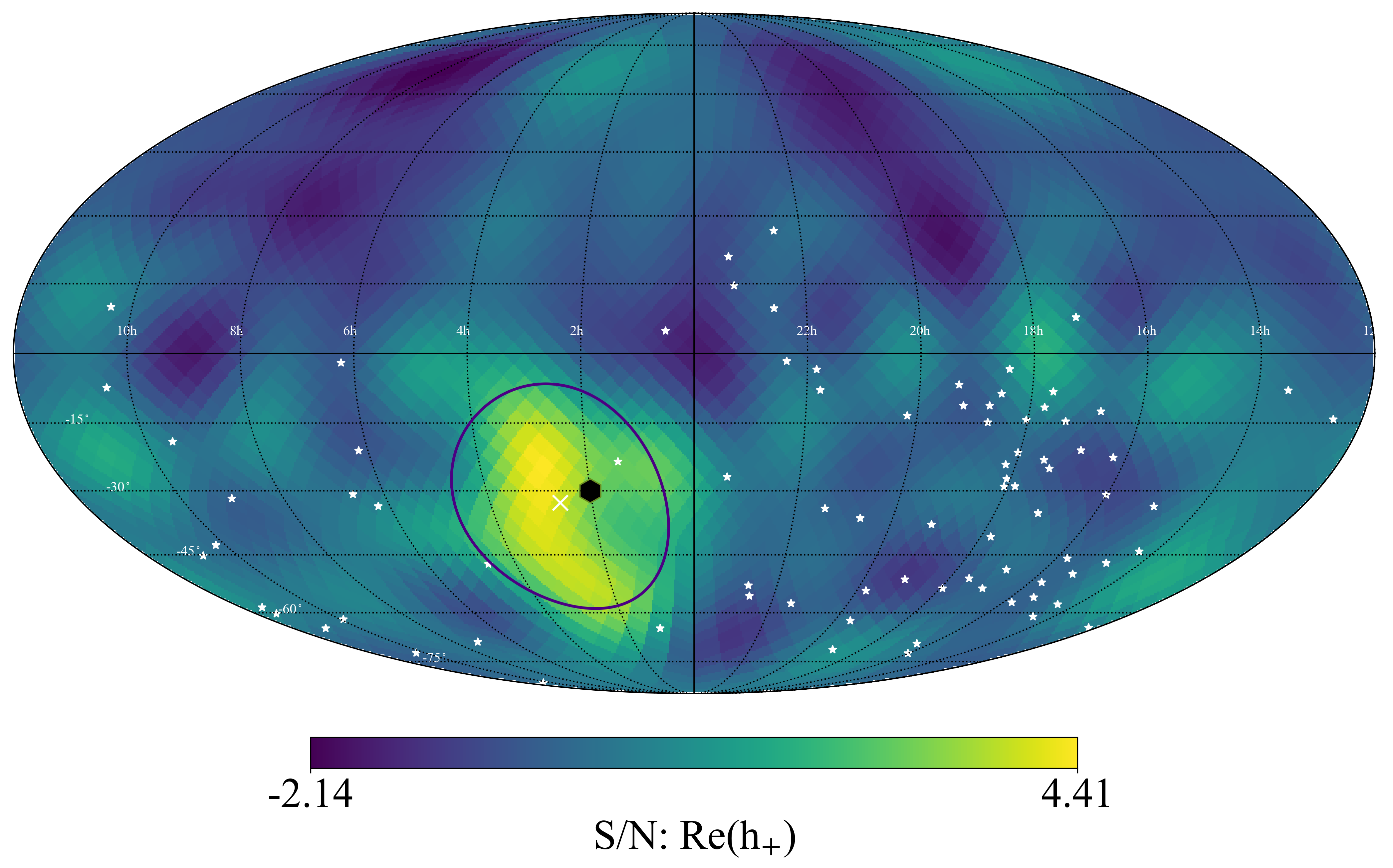}
    \label{fig:bl}
  \end{subfigure}
  \hfill
  \begin{subfigure}[b]{0.48\textwidth}
    \centering
    \includegraphics[width=\textwidth]{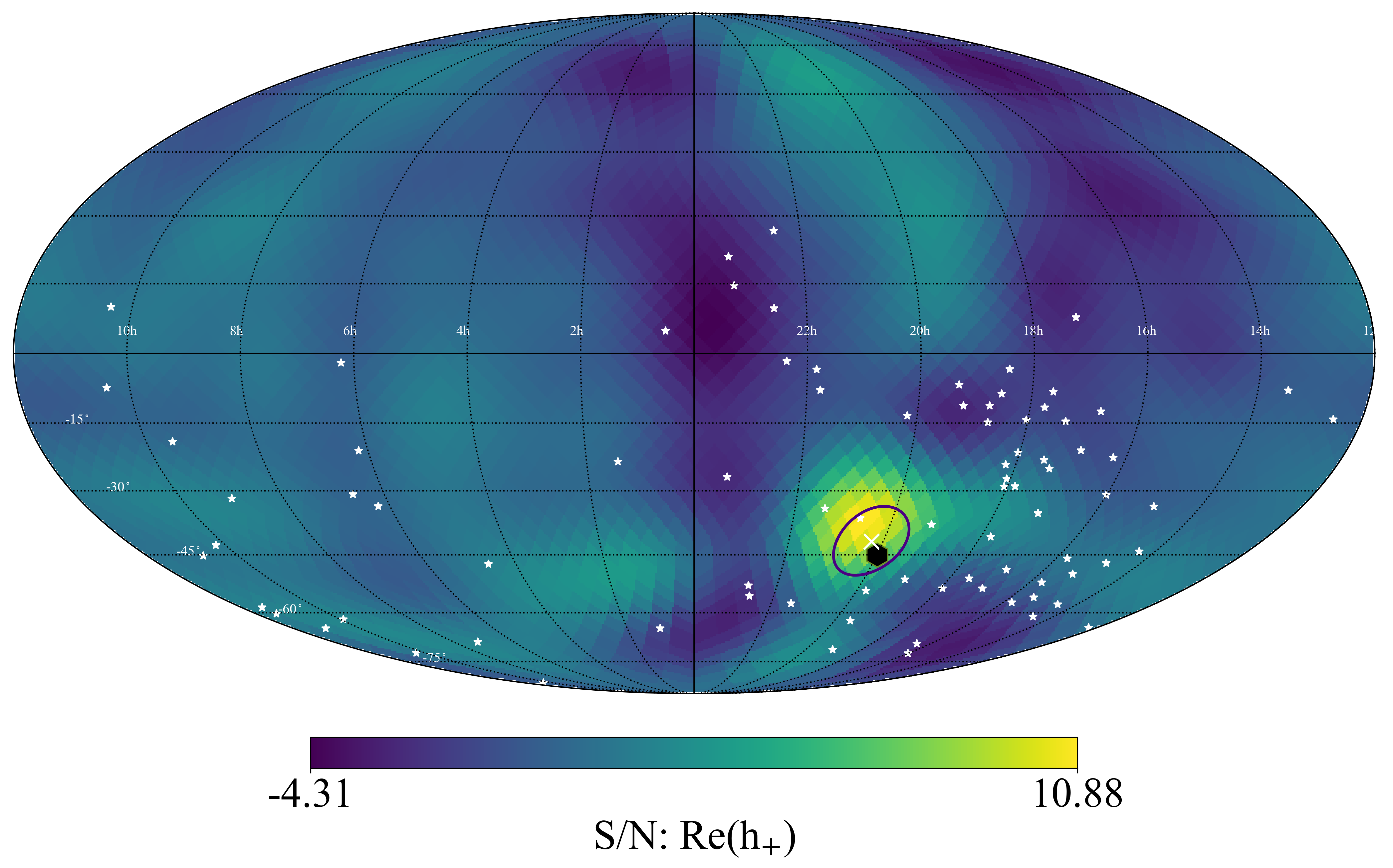}
  \end{subfigure}
  \caption{Clean maps of the real component of $h_+$ polarisation state for the MPTA simulation. Each map shows the same continuous GW source injected in different sky locations, demonstrating differences in directional sensitivity of the array.}
  \label{fig:MK_loc}
\end{figure*}

In the case of a highly non-isotropic PTA array such as the MPTA, the position of the continuous GW source is critical, which we demonstrate in Figure~\ref{fig:MK_loc}. It shows clean maps of 
$\mathrm{Re}(h_+)$ for four different source locations using the MPTA simulation. The direction-dependent sensitivity of the array is immediately apparent. The best-covered location 
(bottom-right), where the majority of pulsars are concentrated, yields the highest clean S/N ($\sim\!10$) with a compact detection region of a single pixel. At the remaining three locations, the S/N drops 
to $\sim\!4$--$6$ and the detection regions expand to multiple pixels, reflecting the broader point spread function in poorly-covered sky directions. Nevertheless, each map shows clear anisotropy consistent with the injected source position. By comparing the clean map S/N at the source pixel to a noise-only simulation, we find that the two locations nearest to the bulk of the array pulsars (bottom-right and upper-right) show significant excess above the noise floor ($>4.5$), indicating reliable detections. At the two antipodal locations (upper-left and bottom-left), the excess is marginal and the apparent signal's amplitude cannot be confidently estimated. 

\begin{figure}
  \centering
  \begin{subfigure}[b]{0.48\textwidth}
    \centering
    \includegraphics[width=\textwidth]{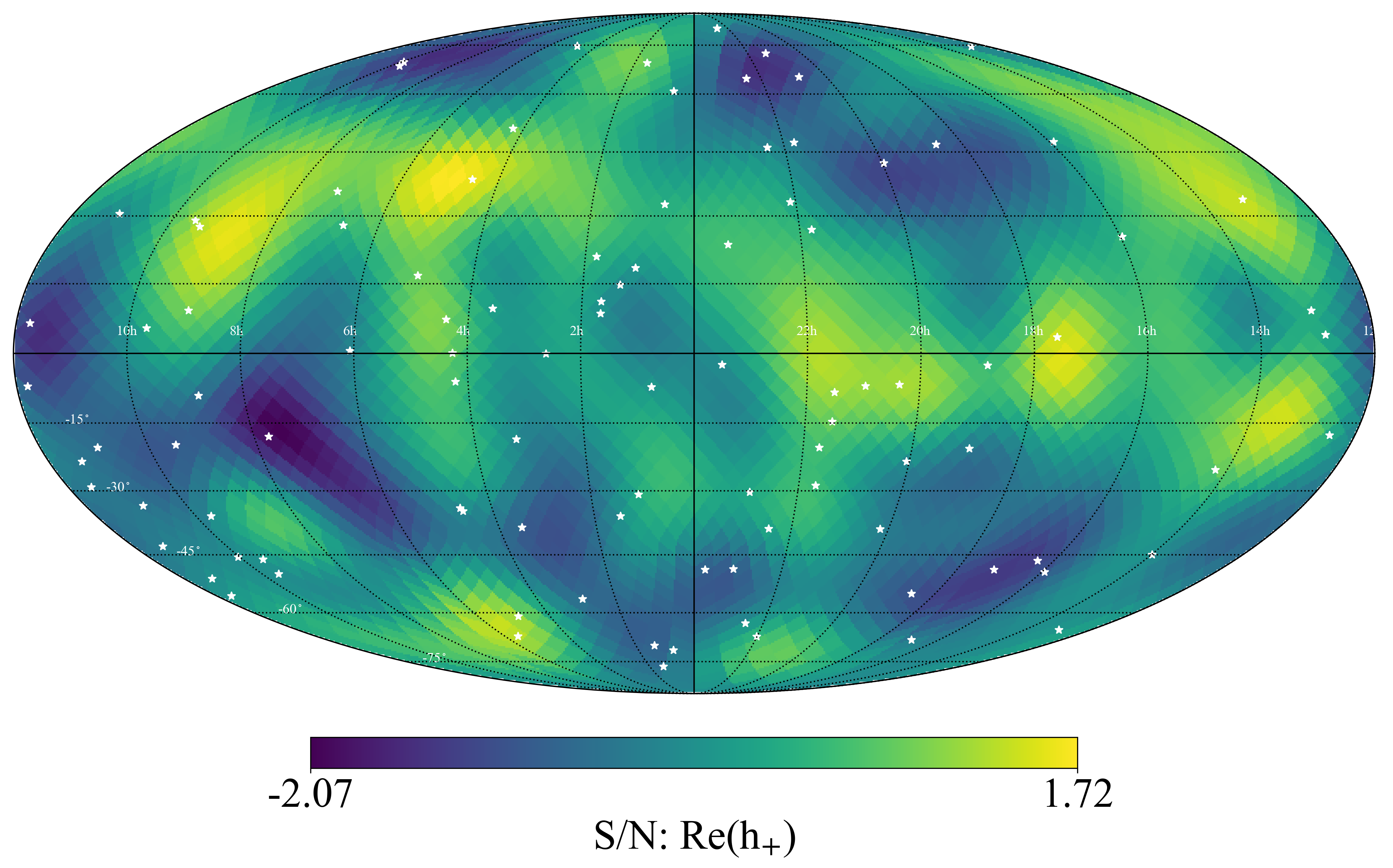}
    \caption{Noise-only simulation (includes white and red noise components).}
  \end{subfigure}
  \hfill
  \begin{subfigure}[b]{0.48\textwidth}
    \centering
    \includegraphics[width=\textwidth]{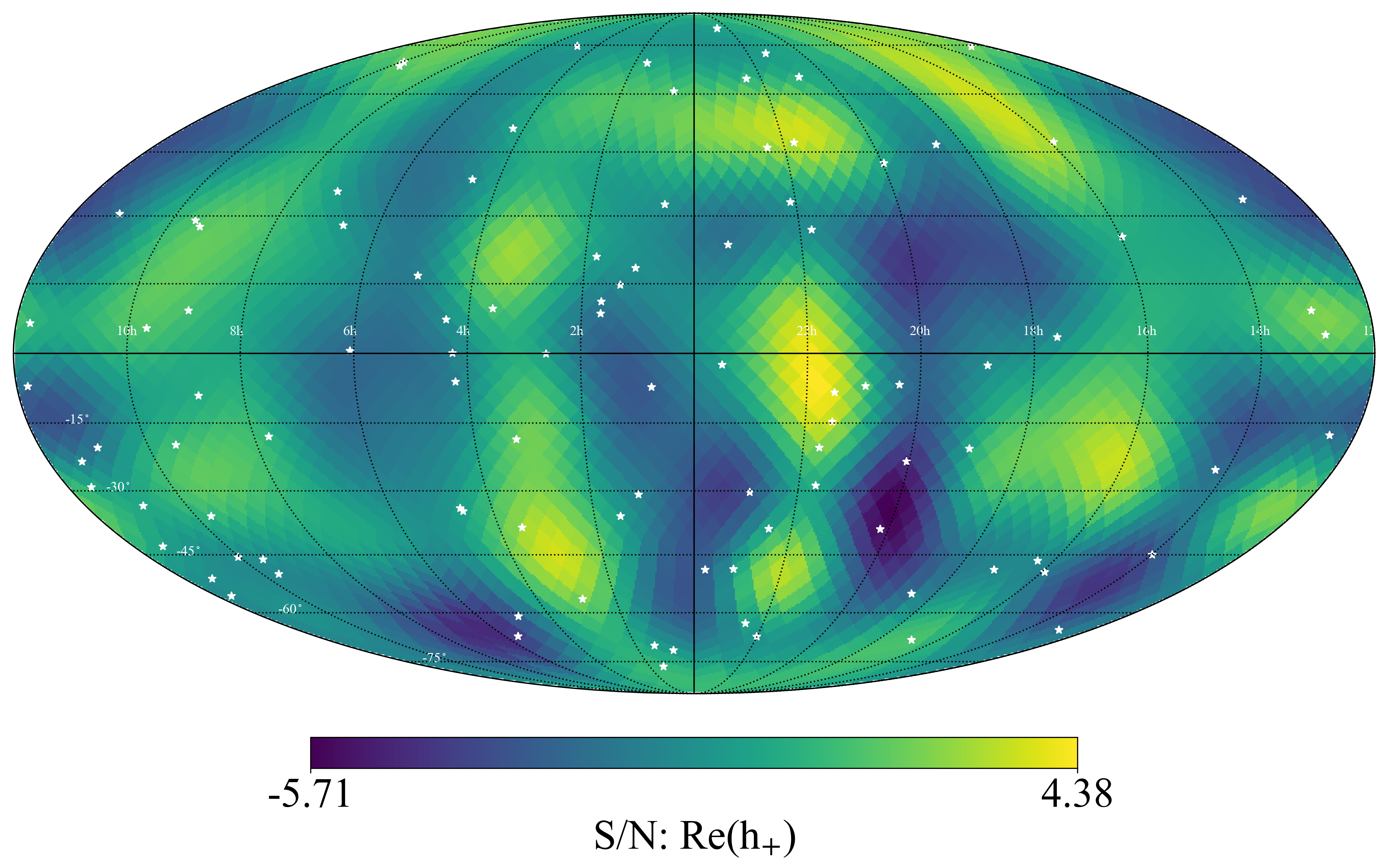}
    \caption{GW background. }
  \end{subfigure}
  \hfill
  \begin{subfigure}[b]{0.48\textwidth}
    \centering
    \includegraphics[width=\textwidth]{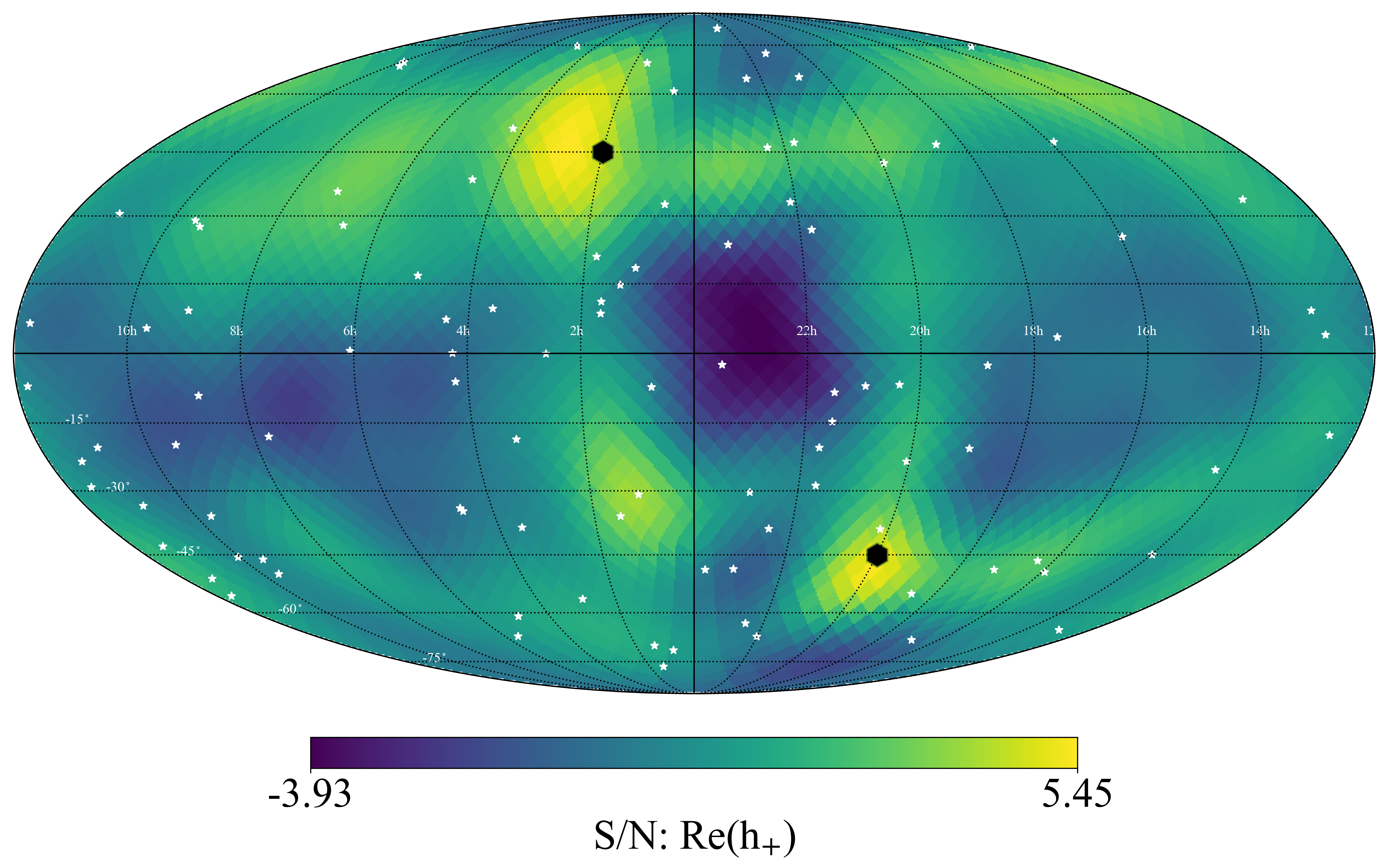}
    \caption{Two continuous GW sources.}
  \end{subfigure}
  \caption{Clean S/N maps of the real $h_+$ polarisation for three types of signals: a) noise only, b) GW background, and c) two identical continuous GW sources.}
  \label{fig:3sigs}
\end{figure}

Finally, Figure~\ref{fig:3sigs} shows three clean maps of $\mathrm{Re}(h_+)$ at the second frequency bin for the isotropic IPTA-like array, illustrating three signal scenarios: noise only, an isotropic GW background, and two continuous GW sources injected with identical parameters.
 
The noise-only and GW background maps both show isotropic structure. In the noise-only map, the S/N is, as expected, of the order of $\sim2$ while the GW map has elevated S/N reflecting the additional power from the stochastic background. The GW background signal is isotropic across all complex component maps and its signal decreases with increasing frequency, as expected for an unpolarised, uncorrelated signal (see Figure~\ref{fig:gwb_pol} in Appendix~\ref{appB}, where we show all complex polarisation maps of the GW background at the first Fourier-frequency bin). 
 
In the two-source map, both continuous GW sources are clearly separated despite sharing identical parameters, including the GW frequency. The recovered S/N of the two sources is comparable, with the source at the better-covered location (bottom-right) showing slightly higher S/N. This demonstrates the ability of our mapping framework to identify and separate multiple sources at the same frequency bin---a capability that will be essential for future PTA observations. If the nanohertz GW signal is dominated by SMBH binaries, the sky is expected to contain multiple overlapping sources within the same frequency bin, and the ability to resolve them individually will be critical for characterising the source population.

\section{Discussion and future work}\label{sec:discussion}

We build on ideas presented in \cite{Gair14,Cornish14} turning them into a universal pipeline for use on PTA data. In this work, we present the basics of the methodology, our specific implementation and test the pipeline's performance using sets of realistic simulations. Below, we broadly discuss our results and further developments which will be included in the next release. 

Our simulation tests demonstrate several key properties of the phase-coherent mapping framework. The radiometer map provides an optimal estimate of the strain amplitude at every sky pixel, while the clean map is essential for source localisation. The highest-S/N detection regions range from a single pixel at well-covered locations to up to ${\sim}\,10$ pixels at poorly-covered directions, reflecting the general resolution limitations of current PTAs rather than a  shortcoming specific to our mapping approach.

We assess the reliability of the radiometer amplitude by comparing the clean map S/N at the source pixel to its value in a noise-only simulation. At sky locations where the clean map excess over noise is significant ($\Delta\mathrm{S/N} \gtrsim 3$), the radiometer recovery of $\mathrm{Re}(h_+)$ is robust (${\sim}\,90$--$97\%$ of the injected value). At locations where the excess is marginal ($\Delta\mathrm{S/N} \lesssim 2$), the radiometer value cannot be reliably separated from the noise fluctuations. The clean map S/N thus serves as a necessary quality criterion for any radiometer-based amplitude measurement. We defer the rigorous statistical assessment of detection significance, in which the observed S/N is compared to a distribution of noise-only realisations, to future work. Our framework naturally supports this procedure, as the noise covariance and Fisher matrices needed to generate synthetic realisations are already computed at each frequency bin.

Most importantly, even when the radiometer map is visually uninformative and the clean map signal is spread across many 
pixels rather than concentrated in a well defined hot spot, the strain amplitude read from the radiometer at the pixel where the continuous GW source was injected still remains consistent with the injected value (provided that the clean map still offers corroborating evidence for a detection). This has direct implications for multi-messenger astronomy: by cross-matching GW sky maps with electromagnetic surveys of SMBH binary candidates, it may be possible to extract reliable GW strain amplitudes at candidate locations even when the GW maps alone cannot localise the source well. The mapping framework thus provides a natural interface between PTA observations and targeted electromagnetic searches, enabling constraints on individual sources that would not be accessible from either data set in isolation.

\subsection{Pulsar term}\label{sec:pulsar_term}
A key avenue for improvement of the continuous GW localisation and parameter estimation is the refinement of pulsar distance measurements. They remain poorly constrained and limit the accuracy of pulsar term modelling to the extent that it is usually treated as an additional source of noise (see e.g., \citealp{Kato26, Grunthal26b}). Even a small $\sim10$\% error in distance estimate distorts the pulsar term given in the bottom of the Eq.~\ref{eq:terms}, by roughly $\Delta\Phi_P \sim \mathcal{O}(1)$~radian, effectively randomizing the phase. In practice, PTAs do not typically model the pulsar term, but marginalise over it. However, if the pulsar distance estimates were improved, the pulsar term could be used to improve the sky localization of continuous GW sources by effectively turning each Earth-pulsar baseline into an independent directional measurement (see e.g., \citealp{Tsai25}).

In our pipeline, the pulsar term can simply be absorbed by the antenna factor \citep{Taylor20}, which can be divided into two components as $\mathcal{F}^{A}_{E}$ and $\mathcal{F}^{A}_{P}$ where
\begin{align}
    \mathcal{F}^{A}_{P} = \mathcal{F}^{A}_{E}e^{-i\Phi_P}
\end{align}
We aim to include pulsar term modelling in a subsequent release of \textsc{MIMOSIS}. The effect of the noise from the unmodelled pulsar term is presented in Fig.~\ref{fig:pterm} in Appendix~\ref{appB}.

\subsection{Covariance between different frequency bins}
When timing residuals are projected onto a Fourier basis 
with frequencies $k/T_{\rm obs}$, this decomposition 
implicitly assumes that the signal is perfectly periodic over $T_{\rm obs}$, which is generally 
not the case. As a result, the power of a signal at a given frequency leaks into neighbouring bins---a consequence of the 
finite observation window acting as a convolution in the frequency domain \citep{windows}. This spectral leakage introduces covariance between frequency bins, which must be treated carefully in PTA analyses \citep{Cristostomi25}. Ignoring these inter-bin correlations can lead to biased inference of GW signal properties, particularly for narrowband signals such as continuous GWs whose frequency may not coincide with a PTA bin frequency $k/T_{\rm obs}$.

The covariance between $n_i(f_k)$ and $n_i(f_l)$---both measured for the same pulsar -- can be written as:
\begin{equation}
\begin{aligned}
    C_{kl} &\equiv \langle n_i^\dagger(f_k) n_i(f_l) \rangle \\
           &= F^\dagger_{ka} C^i_{ab} F_{bl}.
\end{aligned}
\end{equation}
The full covariance matrix across the whole PTA will be then given by:

\begin{equation}
\begin{aligned}
    C_{ijkl} &\equiv \langle n_i^\dagger(f_k) n_j(f_l) \rangle \\
             &= F^\dagger_{ka} C^i_{ab} F_{bl} \, \delta_{ij}.
\end{aligned}
\end{equation}
In practice, the dirty maps in our pipeline remain unchanged, though, the Fisher matrix gains off-diagonal terms reflecting the correlations between different frequency bins. 
These correlations must be taken into account in order to extract, e.g., a continuous wave signal spanning multiple frequency bins.
We will present detailed studies of spectral leakage and a continuous GW parameter estimation in coherently combined maps in an upcoming follow up paper. 

\subsection{Connection to the isotropic analysis}
In the standard isotropic analysis, one assumes that the gravitational-wave background is unpolarized and Gaussian.
It follows that the prior for $h_\nu$ is a Gaussian distribution with mean zero and variance $\sigma^2$ for all values of $\nu$:
\begin{align}
    \pi(h_\nu | \sigma^2) = {\cal N}(h_\nu | \mu=0, \sigma^2) .
\end{align}

If we treat the gravitational-wave strain $h_\nu$ as nuisance parameters, we can marginalize over it to obtain a likelihood for the data conditioned on $\sigma^2$:
\begin{align}
    {\cal L}(\delta t_i | \sigma^2) = & \int dh_\nu \,
    \frac{1}{2\pi \text{det}(C)}  e^{-\frac{1}{2}
    (\delta t_i - \mathcal{F}_{i\nu}h_\nu)^\dagger
    C_{ij}^{-1}
    (\delta t_j - \mathcal{F}_{j\mu}h_\mu)
    } \nonumber\\
    & \pi(h_\nu | \sigma^2) \nonumber\\
    = & \frac{1}{2\pi \det\big(C'(\sigma^2)\big)}
    e^{-\frac{1}{2}\delta t_i^\dagger C'_{ij}(\sigma^2) \delta t_j} .
\end{align}
The new likelihood (familiar to pulsar timers) is still Gaussian, but now with mean zero---and the covariance matrix has gained a gravitational-wave component, which depends on the strength of the stochastic background $\sigma^2$.
Thus, the isotropic likelihood is a special case of the phase-coherent likelihood that can be obtained by assuming that the elements of the gravitational-wave sky map are drawn from an isotropic, unpolarized, Gaussian distribution.

\section{Conclusions}\label{sec:conclusions}
We present a comprehensive framework to study low-frequency GWs with phase-coherent maps. It applies a universal approach to analyse all kinds of signals such as stochastic backgrounds, anisotropy signatures and individual sources. It is also fully compatible with established pulsar timing data analysis methods. 

In this work we describe the backbone of the framework, which we call \textsc{MIMOSIS}, and test it with a set of simulations imitating current PTA capabilities. In follow-up studies we aim: to study the statistical properties of signals that can be probed with phase-coherent mapping, to quantify anisotropy, and to recover binary parameters. 
This will be followed by a public release of the code. Finally, we aim to apply the formalism to analyse the real data.

The main results and conclusions of our work can be summarised as follows:
\begin{itemize}
     \item We can successfully recover anisotropy for a wide range of scenarios including different distributions of pulsars, different noise properties, and different injected sources. 

    \item Complex polarisation maps contain information useful for single GW source analysis, which can be used to constrain single binary parameters.

    \item The radiometer and clean maps play complementary roles. The radiometer may provide unbiased strain amplitude estimates, while the clean map---obtained via regularised inversion of the full Fisher matrix---is essential for source localisation. At sky locations where the clean map S/N significantly exceeds the noise floor, the radiometer amplitude recovery is robust (${\sim}\,90$--$97\%$). This allows for the targeted strain measurements at known SMBH binary positions from electromagnetic catalogues.

    \item The key advantage of phase-coherent mapping is its capability to robustly combine information from different maps at multiple frequencies, thereby boosting the significance of CW signals spread over multiple bins.

    \item As noted in \cite{Gair14}, phase-coherent maps can be reduced to reproduce maps of cross-correlated power, which makes the two approaches translatable and complementary. Future practical implementation of these ideas in \textsc{MIMOSIS} may therefore serve as a versatile analysis tool. 

\end{itemize}

\begin{acknowledgments}
The authors are supported via the Australian Research Council (ARC) Centre of Excellence CE230100016.
This work made use of \texttt{libstempo} \citep{Vallisneri20} for 
pulsar timing simulations and signal injection, and  \textsc{Enterprise} \citep{enterprise} for noise model definitions 
and signal parameterisations. Sky maps were constructed using the 
\textsc{HEALPix} pixelisation scheme \citep{Gorski05} via the 
\textsc{healpy} package. Numerical computations relied on \textsc{NumPy} and \textsc{SciPy}, and figures were produced with \textsc{Matplotlib}.

\end{acknowledgments}

\appendix

\section{Additional equations}\label{appA}
This appendix collects additional equations underlying the methodology presented in this work. We define the geometric and signal model quantities required to describe the GW-induced timing residuals in a PTA, including the antenna pattern functions, the GW polarization basis vectors, the Fourier design matrix, and the time domain signal model for a continuous GW from an SMBH binary. 

The antenna pattern functions $\mathcal{F}^+$ and $\mathcal{F}^\times$ describe the 
sensitivity of a given pulsar to each GW polarization, and depend on the geometry 
between the pulsar direction $\hat{p}$ and the GW propagation direction $\hat{\Omega}$ \citep{Sesana10},
\begin{equation}
  \begin{aligned}
    \mathcal{F}^+(\hat{p},\hat{\Omega}) =& \frac{1}{2}  \frac{\left( \hat{m} \cdot \hat{p} \right)^2 - \left( \hat{n} \cdot \hat{p} \right)^2}{1 + \hat{\Omega} \cdot \hat{p}} \\ \\
    \mathcal{F}^\times(\hat{p},\hat{\Omega}) =& \frac{\left( \hat{m} \cdot \hat{p} \right) \left( \hat{n} \cdot \hat{p} \right)}{1 + \hat{\Omega} \cdot \hat{p}}
\end{aligned}  
\end{equation}
where $\hat{m}$ and $\hat{n}$ are the GW polarization basis vectors, defined in terms 
of the source sky position $(\theta, \phi)$ and polarization angle $\psi$ as
\begin{equation}
\begin{aligned}
\vec{m} &=
\left(\sin\phi\cos\psi - \sin\psi\cos\phi\cos\theta\right)\hat{x}
-\left(\cos\phi\cos\psi + \sin\psi\sin\phi\cos\theta\right)\hat{y}
+\left(\sin\psi\sin\theta\right)\hat{z}, \\[6pt]
\vec{n} &=
\left(-\sin\phi\sin\psi - \cos\psi\cos\phi\cos\theta\right)\hat{x}
+\left(\cos\phi\sin\psi - \cos\psi\sin\phi\cos\theta\right)\hat{y}
+\left(\cos\psi\sin\theta\right)\hat{z}.
\end{aligned}
\end{equation}
The GW propagation direction is defined as $\hat{\Omega} = \vec{m}\times\vec{n}$,
\begin{equation}
\hat{\Omega}
= -(\sin\theta\cos\phi)\,\hat{x}
-(\sin\theta\sin\phi)\,\hat{y}
-\cos\theta\,\hat{z}.
\end{equation}
Note that the direction \textit{towards} the GW source is denoted with $\hat{k} = -\hat{\Omega}$.

The sky position of pulsar $i$ is described by polar coordinates 
$(\theta_i, \phi_i)$, encoded in the unit vector
\begin{equation}
\hat{p}_{i}
= (\sin\theta_{i}\cos\phi_{i})\,\hat{x}
+(\sin\theta_{i}\sin\phi_{i})\,\hat{y}
+\cos\theta_{i}\,\hat{z}.
\end{equation}
The timing residuals are projected onto a Fourier basis via the design matrix 
$\mathbf{F}$ \citep{Taylor21}, whose columns consist of sine and cosine functions 
at harmonics $k/T_{\rm obs}$ for $k = 1, \dots, N_f$, evaluated at each TOA $t_m$ for $m= 1,...,N_m$:
\[
\mathbf{F} =
\begin{pmatrix}
\sin\!\left(\frac{2\pi t_1}{T}\right) & \cos\!\left(\frac{2\pi t_1}{T}\right) & \cdots &
\sin\!\left(\frac{2\pi N_f t_1}{T}\right) & \cos\!\left(\frac{2\pi N_f t_1}{T}\right) \\
\sin\!\left(\frac{2\pi t_2}{T}\right) & \cos\!\left(\frac{2\pi t_2}{T}\right) & \cdots &
\sin\!\left(\frac{2\pi N_f t_2}{T}\right) & \cos\!\left(\frac{2\pi N_f t_2}{T}\right) \\
\vdots & \vdots & \ddots & \vdots & \vdots \\
\sin\!\left(\frac{2\pi t_{N_m}}{T}\right) & \cos\!\left(\frac{2\pi t_{N_m}}{T}\right) & \cdots &
\sin\!\left(\frac{2\pi N_f t_{N_m}}{T}\right) & \cos\!\left(\frac{2\pi N_f t_{N_m}}{T}\right)
\end{pmatrix}
\]
\noindent The GW-induced residuals in time-domain are given by
\begin{align}\label{eq:appAstart}
r_+(t) &= \alpha(t)\left[A(t)\cos 2\psi + B(t)\sin 2\psi\right], \\
r_\times(t) &= \alpha(t)\left[-A(t)\sin 2\psi + B(t)\cos 2\psi\right],
\end{align}
where the amplitude, and the $\sin$/$\cos$ coefficients are
\begin{align}
\alpha(t) &= \frac{\mathcal{M}^{5/3}}{ D_L\, \omega(t)^{1/3}}, \\
A(t) &= \frac{1}{2}(3 + \cos 2\iota)\sin 2\Phi(t), \\
B(t) &= 2\cos\iota \cos 2\Phi(t),
\end{align}
and $\psi$ is the polarization angle, $\iota$ the inclination, and $D_L$ the 
luminosity distance. The evolving orbital frequency and phase are
\begin{align}\label{eq:appAend}
\omega(t) &= \omega_0\left(1 - \frac{256}{5}\mathcal{M}^{5/3}\omega_0^{8/3}\,t\right)^{-3/8}, \\
\Phi(t) &= \Phi_0 + \frac{1}{32\mathcal{M}^{5/3}}\left(\omega_0^{-5/3} - \omega(t)^{-5/3}\right),
\end{align}
where $\omega_0 = \pi f_{\rm gw}$ is the initial orbital angular frequency and 
$\Phi_0 = \phi_0/2$ is the initial orbital phase and $\phi_0$ is the initial GW phase. 

\section{Maps of the GW background and reconstructions affected by pulsar term noise}\label{appB}
\begin{figure}
  \centering
  \includegraphics[width=\columnwidth]{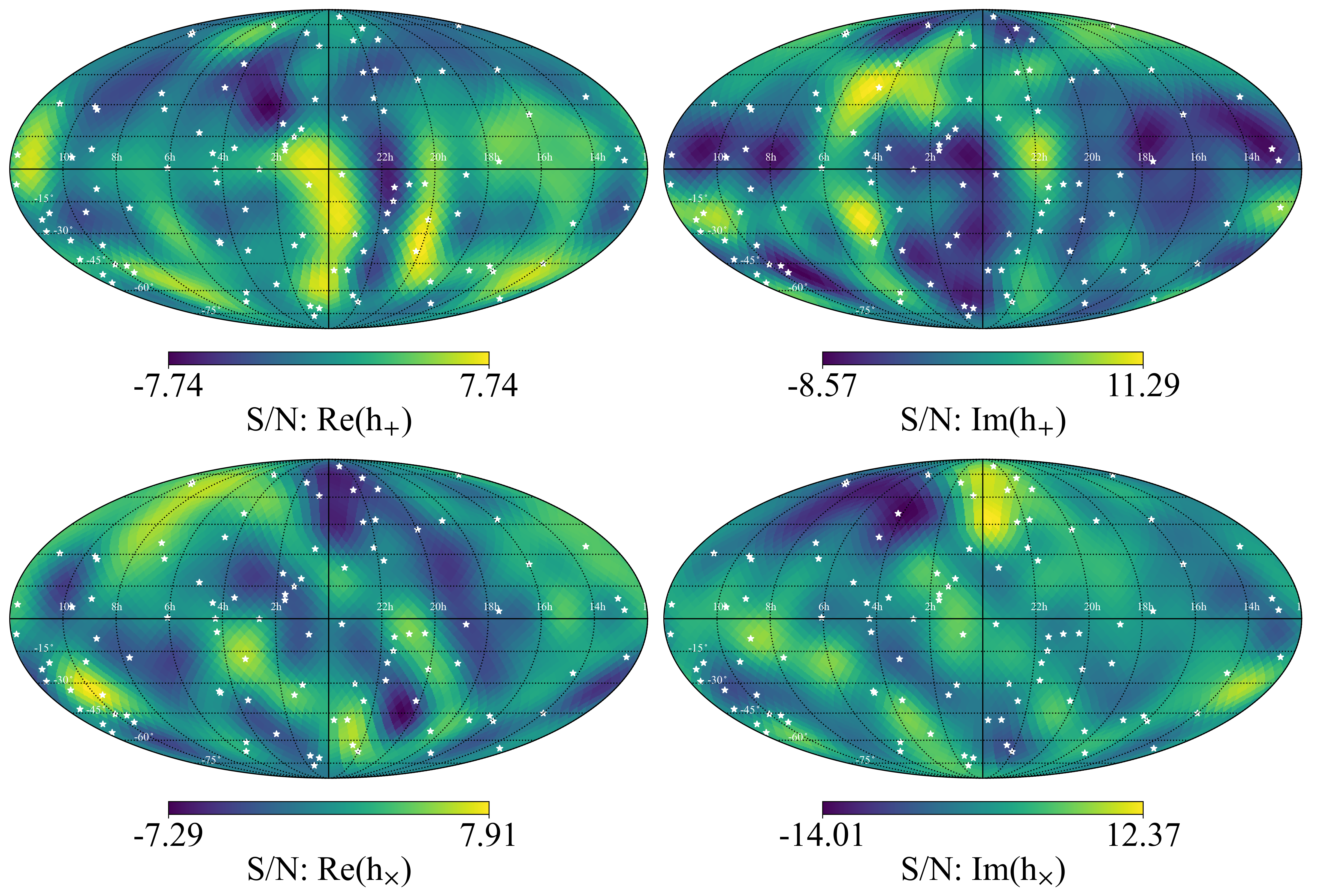}
  \caption{Clean S/N maps for an isotropic GW background at the first Fourier-frequency bin for the IPTA simulation.}
  \label{fig:gwb_pol}
\end{figure}

Figure~\ref{fig:gwb_pol} shows the clean map S/N in all four 
components of the complex polarisations for an isotropic GW background at the first Fourier-frequency bin. All four maps show comparable S/N levels with no preferred sky direction or polarisation, consistent with an unpolarised, uncorrelated and isotropic stochastic background. The elevated S/N at this frequency bin compared to higher bins (shown in the main text) reflects the steep red power-law spectrum of the injected background.

\begin{figure}
  \centering
  \begin{subfigure}[b]{\columnwidth}
    \centering
    \includegraphics[width=0.9\textwidth]{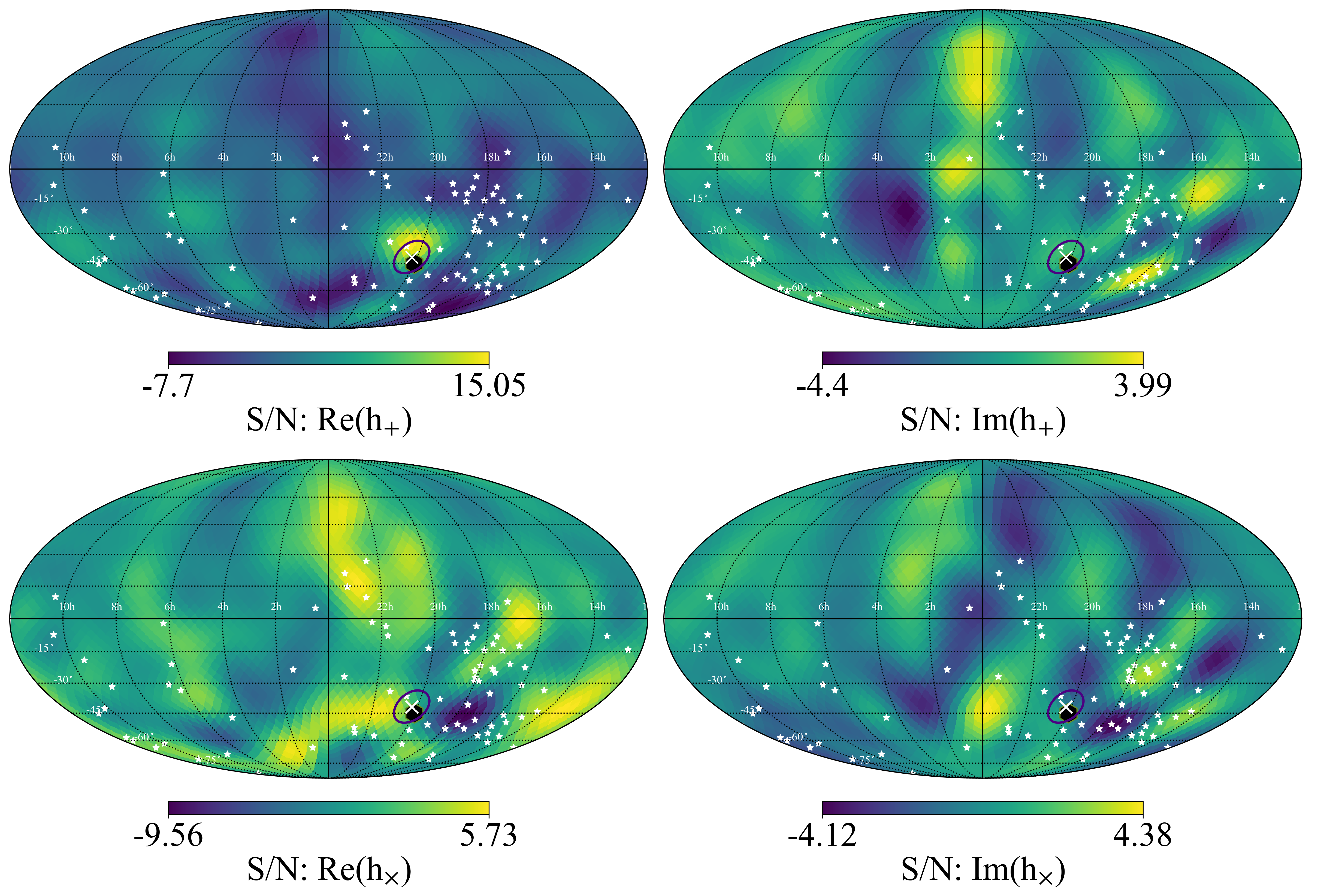}
  \end{subfigure}
  \vspace{0.3cm}
  \begin{subfigure}[b]{\columnwidth}
    \centering
    \includegraphics[width=0.9\textwidth]{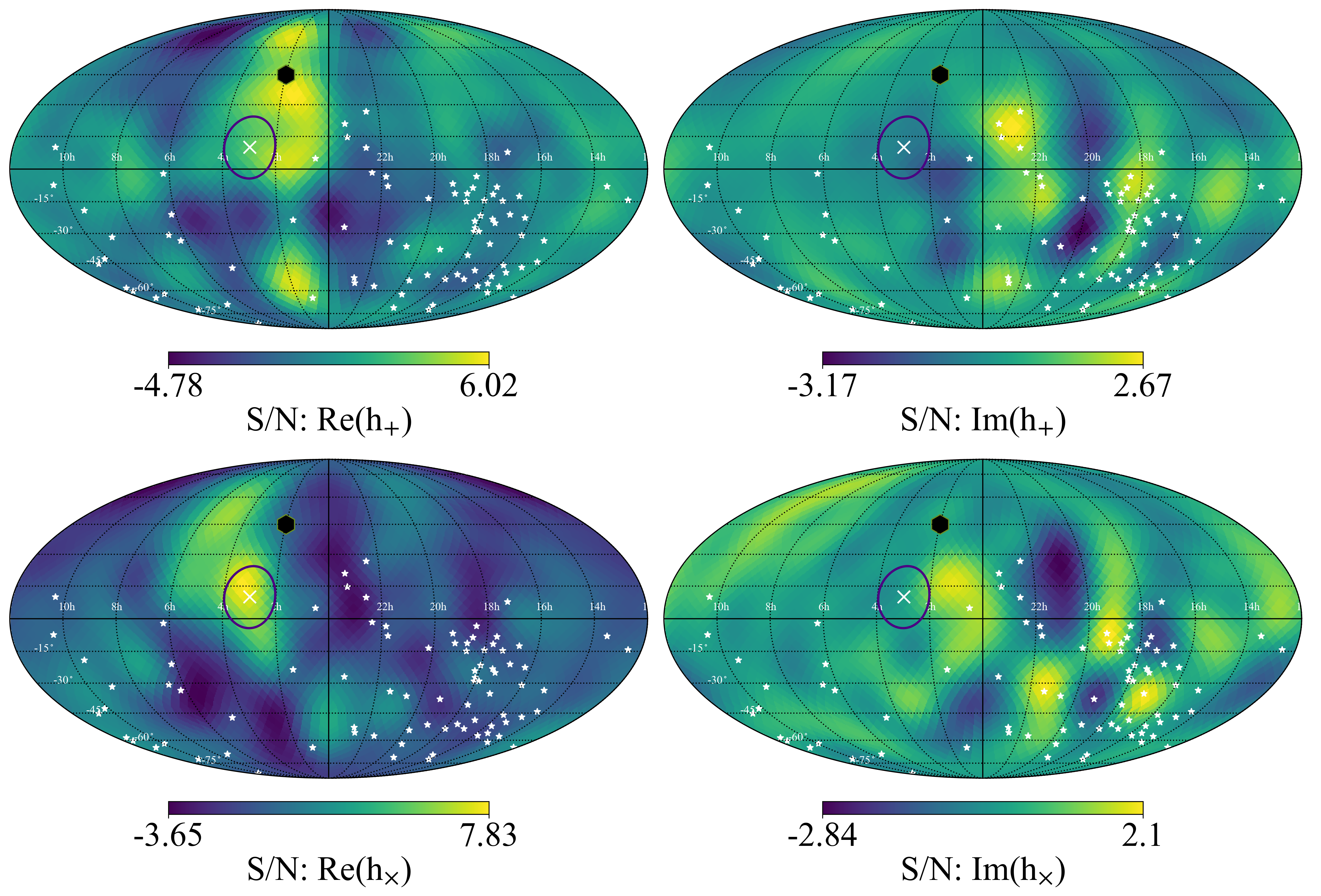}
  \end{subfigure}
  \caption{Impact of the unmodelled pulsar term on the clean map S/N for two sky locations in the MPTA-like array. Top: best-covered location (bottom-right), where the coherent earth term dominates and the source is correctly recovered. Bottom: poorly-covered location (upper-left), where the incoherent pulsar term power is comparable to the Earth-term signal, degrading both localisation and amplitude recovery.}
  \label{fig:pterm}
\end{figure}

Figure~\ref{fig:pterm} shows the impact of including the pulsar term in the injected continuous GW signal while modelling only the earth term in the mapping framework. We present clean map S/N in all four complex polarisation channels for the MPTA-like array at two sky locations: the best-covered direction (bottom-right, panel a) and a poorly covered direction (upper-left, panel b). In both cases, the injected signal is pure $\mathrm{Re}(h_+)$, so any power in the remaining three channels is attributable to noise including the unmodelled pulsar term.

At the best-covered location (bottom-right), the Earth-term signal dominates: the $\mathrm{Re}(h_+)$ S/N at the source pixel is $\sim 35$ (at the radiometer map), and the source is correctly localised. However, the unmodelled pulsar term introduces additional GW power that is incoherently distributed across all four complex maps, elevating the S/N above noise levels. 
 
At a poorly-covered location (upper-left), the Earth-term S/N ($\sim\!6$) is comparable to the pulsar term contribution, which scrambles the signal phase and prevents coherent recovery.
 
These results demonstrate that the Earth-term approximation is valid where the array has good sensitivity---the coherent earth term dominates over the incoherent pulsar term. At poorly-covered locations, the pulsar term can dominate and degrade both localisation and amplitude recovery. For arrays with non-isotropic pulsar distributions, the reliability of the Earth-term approximation is therefore direction-dependent.

\bibliography{main}{}
\bibliographystyle{aasjournalv7}
\end{document}